\documentclass[english]{FCFM-UAS-reporte}
\usepackage{achicago,boxedminipage}
\usepackage{caption}
\begin{document}

% Portada

\begin{titlepage}
\begin{center}
\setlength{\fboxrule}{0.1cm} 
\setlength{\fboxsep}{1cm} 
%\begin{boxedminipage}[21.94cm][c]{15.59cm}
\framebox[15.59cm]{
\begin{minipage}[21.94cm]{15.59cm}
\begin{center}
\parbox{13.59cm}{
\begin{center}
\renewcommand{\baselinestretch}{1.5}
{\small\Large\textbf{Combining Cluster Sampling and Link-Tracing Sampling to Estimate Totals and
		Means of Hidden Populations in Presence of Heterogeneous Probabilities of Links}}
\normalsize
\end{center}
}

\vspace{1.0cm}

Mart{\'\i}n H. F\'elix Medina

\vspace{2.0cm}
 Reporte T\'ecnico No. FCFM-UAS-2020-01
 
\vspace{0.5cm}
 Serie: Investigaci\'on
 
\vspace{0.5cm}
 21 de Mayo de 2020
 
\vspace{0.5cm}

\includegraphics[width=3.0cm,height=3.5cm]{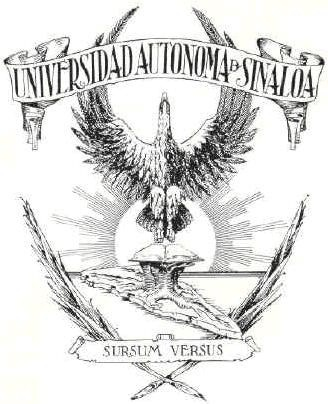}
  
\vspace{0.5cm}
 Facultad de Ciencias F{\'\i}sico-Matem\'aticas

\vspace{0.5cm}
 Universidad Aut\'onoma de Sinaloa
 
\vspace{0.5cm}
Ciudad Universitaria, Culiac\'an Sinaloa

\vspace{0.5cm}
M\'exico

\end{center}

\end{minipage}

}

\end{center}
%\end{boxedminipage}

\end{titlepage}

% Primera pagina

\title{{\Large\textbf{Combining Cluster Sampling and Link-Tracing Sampling to Estimate Totals and
			Means of Hidden Populations in Presence of Heterogeneous Probabilities of Links}}}

\author{Mart{\'\i}n H. F\'elix Medina\thanks{mhfelix@uas.edu.mx}\\Facultad de Ciencias 
 F{\'\i}sico-Matem\'aticas de la\\Universidad Aut\'onoma de Sinaloa}

\date{}

\maketitle

\vspace{-1ex}

\begin{abstract}
We propose Horvitz-Thompson-like and H\'ajek-like estimators of the total and mean of the values of a variable 
of interest associated with the elements of a hard-to-reach population sampled by the variant of link-tracing 
sampling proposed by F\'elix-Medina and Thompson (2004). As examples of this type of population are drug users, 
homeless people and sex workers. In this sampling variant, a frame of venues or places where the members of the 
population tend to gather, such as parks and bars, is constructed. The frame is not assumed to cover the whole 
population. An initial cluster sample of elements is selected from the frame, where the clusters are the venues, 
and the elements in the initial sample are asked to name their contacts who are also members of the population. 
The sample size is increased by including in the sample the named elements who are not in the initial sample.
The proposed estimators do not use design-based inclusion probabilities, but model-based inclusion probabilities 
which are derived from a model proposed by F\'elix-Medina et al. (2015) and are estimated by maximum likelihood 
estimators. The inclusion probabilities are assumed to be heterogeneous, that is, that they depend on the sampled 
people. Estimates of the variances of the proposed estimators are obtained by bootstrap and they are used to 
construct confidence intervals of the totals and means. The performance of the proposed estimators and confidence 
intervals is evaluated by two numerical studies, one of them based on real data, and the results show that their 
performance is acceptable.
\end{abstract}

\vspace{1ex}
 
\sloppypar{\textbf{Key words}: \textit{Chain-referral sampling, capture-recapture, H\'ajek estimator, Horvitz-Thompson estimator, 
	   maximum likelihood estimator, snowball sampling}}

\newpage

\begin{center}
\textbf{\small{Resumen}}
\end{center}
{\small En este trabajo se proponen estimadores tipo Horvitz-Thompson y tipo H\'ajek del total y la media de los valores 
	de una variable de inter\'es asociados con los elementos de una poblaci\'on de dif{\'\i}cil detecci\'on que se muestrea 
	mediante la variante del muestreo por bola de nieve propuesta por F\'elix-Medina y Thompson (2004). Ejemplos de este 
	tipo de poblaci\'on son drogadictos, indigentes y trabajadoras sexuales. En esta variante de muestreo se construye un 
	marco muestral de sitios donde los miembros de la poblaci\'on tienden a reunirse, tales como parques y bares. No se 
	supone que el marco muestral cubre a toda la poblaci\'on. Del marco muestral se selecciona una muestra inicial por 
	conglomerados de elementos, donde los conglomerados son los sitios, y se les pide a los elementos en la muestra inicial 
	que nombren a sus contactos que tambi\'en sean miembros	de la poblaci\'on. El tama\~no muestral se incrementa al incluir 
	en la muestra a los elementos nombrados que no est\'an en la muestra inicial. Los estimadores que se proponen no usan 
	probababilidades de inclusi\'on basadas en el dise\~no, sino basadas en un modelo propuesto por F\'elix-Medina et al. 
	(2015) y se estiman mediante estimadores m\'aximo veros{\'\i}miles. Las probabilidades de inclusi\'on se consideran que 
	son heterog\'eneas, esto es, que dependen de las personas muestreadas. Estimaciones de las varianzas de los estimadores 
	propuestos se obtienen mediante bootstrap y se usan para construir intervalos de confianza de los totales y medias. Los 
	desempe\~nos de los estimadores e intervalos de confianza propuestos se eval\'uan mediante dos estudios num\'ericos, uno 
	de ellos basado en datos reales, y los resultados muestran que sus desempe\~nos son aceptables.
}

\vspace{2ex}

\section{Introduction}

The problem of selecting samples from hidden or hard-to-detect populations, such as drug users, sex workers and
homeless people, that allow reasonably valid statistical inferences is challenging because of the following factors:
(i) lack of appropriate sampling frames for those populations; (ii) rareness of those populations; (iii) elusiveness 
of their members to be sampled; (iv) difficulty in identifying their elements due to a stigmatized or illegal behaviour; 
(v) difficulty in locating their members, and (v) hardness in persuading their elements to participate in the study, 
among others. See Tourangean (2014) for a detailed discussion about these and other issues. Because of these factors,
conventional sampling methods are not appropriate for this type of population, and consequently several specially
tailored sampling methods that take into account the particular characteristics of those populations have been proposed.
Among these methods we can mention multiplicity sampling, venue-based sampling, link-tracing sampling and capture-recapture
sampling. See Spreen (1992), Magnani et al. (2005), Kalton (2009), Marpsat and Razafindratsima (2010), UNAIDS/WHO Working
Group on Global HIV/AIDS and STI Surveillance (2010), Lee et al. (2014), Spreen and Bogaerts (2015) and Heckathorn and
Cameron (2017) for descriptions of these methods. In addition, special estimation methods based on conventional samples,
such as the scale up method (Killworth et al. 1998a and 1998b, Bernard et al. 2010, McCormick et al. 2010 and Maltiel et al.
2015) or on a combination of some of the previously mentioned sampling designs, such as the multiplier method (UNAIDS/WHO 
Working Group on Global HIV/AIDS and STI Surveillance 2010, Johnston et al. 2013 and Men and Gustafson 2017) and the one 
step network based method (Dombrowski et al. 2012 and Khan et al. 2018) have been proposed.

It is worth noting that most of the recently published research papers on sampling from hidden populations focus on estimating
the population size. (See Cheng et al. 2020 for a review and analysis, from an asymptotic approach, of most of these methods.) 
Thus, the scale up, multiplier and one step network methods were developed with this goal in mind. The interest in developing 
methods for estimating the size of a hidden population is mainly because information about this parameter allows the design of 
appropriate plans to address the problems associated with this type of population. However, information about other population 
parameters, such as average monthly spending on drugs and average age at which consumption begins in a population of drug 
addicts, and average weekly number of clients and average weekly income in a population of sex workers, is also important because 
it increases our knowledge about the population, and in addition, this knowledge could be used to improve the plans for its care 
that are based only on its population size.

On the other hand, among the sampling designs for hidden populations that allow estimating parameters different from or in addition 
to the population size, we have venue-based sampling and link-tracing sampling. Venue-based sampling (MacKellar et al., 1996) is a 
probability sampling method specifically developed to estimate the means of variables of interest, and particularly proportions. 
The method consists in carrying an ethnographic study to construct a sampling frame of venues where the members of the population 
tend to gather. Venues are not only sites, such as bars, parks and street locations, but could be combinations of sites, days of 
the week and time segments. For instance, a venue could be a specific bar from 4:00 p.m. to 12:00 a.m. on Fridays and Saturdays, 
whereas another venue could be the same bar, over the same time segment, but from Monday to Thursday. Furthermore, some venues 
could be events such as gay parades. A probability sample of venues is selected, and from each chosen venue a sort of systematic 
sample of members of the population who are present at the venue is selected. For each sampled element the values of the variables 
of interest associated with that person are recorded, and in addition, information about his or her attendance to the venues in 
the sampling frame is obtained so that his or her inclusion probability can be estimated, and consequently, the estimators of the 
means can be computed. It is evident that the estimates obtained by using this sampling design are valid only for the portion of 
the population that attends the venues in the frame. Therefore, the extension of the results to the entire population requires 
the assumption that with respect to the distributions of the variables of interest there are no differences between the members 
who visit the places in the frame and those who do not.

Link-tracing sampling (LTS) is an umbrella term that encompasses a set of sampling designs in which an initial sample of elements from 
the target population is selected and every sampled person is asked to name his or her contacts (defined according to a certain 
criterion) who are also members of the target population. The elements in the initial sample form the wave zero, and the named elements
who are not in the initial sample form the wave one. People in wave one might also be asked to name their contacts. The named elements
who have not been previously sampled form the wave two. The sampling procedure might continue in this way until a stopping rule is 
satisfied. For instance, a specified number of waves or a specified sample size. Several variants of LTS have been proposed. For example,
Klovdahl (1989) developed a variant, known as random walk, in which the size of each wave is equal to one. Heckathorn (1997 and 2002) 
proposed a variant of LTS, called respondent driven sampling (RDS), in which every person who is already included in the sample is
asked to name a small fixed number of his or her contacts randomly selected. RDS was originally proposed to estimate proportions of some
subpopulations of the population of interest, such as the proportions of women and people under 18 years of age in a population of drug 
users. However, subsequently, Volz and Heckathorn (2008) developed estimators of the total and mean of a quantitative response variable 
and Handcock et al. (2014), as well as Crawford et al. (2018), developed Bayesian models which allow inferences about the population size. 
It is worth noting that RDS is a sampling design that has been extensively used worldwide to select samples from different hard-to-reach 
populations and according to Johnston et al. (2016), it has the endorsement of different organizations such as the US Centers for Disease 
Control and Prevention, UNAIDS, WHO and others. Another variant of LTS is the one proposed by Frank and Snijders (1994). In this variant, 
the initial sample is assumed to be selected by Bernoulli sampling, that is, every element of the population has the same probability of 
being included in the initial sample and the inclusions are independent. Furthermore, those authors assumed that the probability that a 
specific element of the population be named as a contact by a particular person in the initial sample, which in this document is called 
link probability, is a constant, that is, it does not depend on the named person nor on the person who names. This supposition is known 
as the homogeneity assumption. Finally, it is worth noting that Thompson and Frank (2000) proposed a stochastic block model to estimate 
population proportions from samples selected by a pretty general LTS design. They assumed that the population is divided into two 
subpopulations and that the probability of a link between two elements of the population depends on the subpopulations to which those 
elements belong. From their assumed model they construct maximum likelihood estimators of the proportions of those subpopulations. Chow 
and Thompson (2003) used the previous model and derived estimators from a Bayesian approach. Finally, St. Clair and O'Connell (2012) 
extended the previous Bayesian model to estimate the population mean of a quantitative variable of interest. 

In this work we consider the problem of estimating the total and the mean of a variable of interest, such as the weekly drug expense of 
a drug user, the number of weekly clients of a sex worker, and an indicator (positive=1/negative=0) of a person's drug use, from a sample
selected by the LTS variant proposed by F\'elix-Medina and Thompson (2004). This sampling variant was devised to avoid the assumption of 
an initial Bernoulli sample required by the sampling variant proposed by Frank and Snijders (1994), which is difficult to satisfy in 
real-world applications. To achieve this goal, those authors proposed that a sampling frame of venues or places where the members of the
population tend to gather be constructed, as in venue-based sampling. The frame is not assumed to cover the whole population, but only
a portion of it. Then, a simple random sample without replacement (SRSWOR) of venues is selected and the members of the population who 
belong to each sampled venue are sampled. Next, from each sampled venue its elements are asked to named their contacts who are also 
members of the population, either they belong or not to the portion covered by the frame. It is important to indicate that, as in 
venue-based sampling, the venues in the frame could be combinations of places and time segments. However, in this LTS variant, an element 
in the portion covered by the frame is assumed that can be assigned to only one venue by using a specified criterion, for instance, the 
venue where the person spends most of his or her time. F\'elix-Medina and Thompson (2004) proposed maximum likelihood estimators (MLEs)
of the population size which were derived under the assumption that the probability that a person is linked to a sampled venue, that is, 
that he or she is a contact of an element in that venue, depends on the venue, but not on the named person. This means that the estimators 
were derived under the assumption of homogeneous link probabilities. Later, F\'elix-Medina et al. (2015) derived MLEs of the population 
size under the assumption that the link probabilities also depend on the named persons, that is, that the probabilities are heterogeneous.
In this work, we used the model proposed by these authors to construct model-based Horvitz-Thompson-like estimators (HTLEs) and 
model-based H\'ajek-like estimators (HKLEs) of the total and the mean of a variable of interest. It should be noted that F\'elix-Medina 
and Monjardin (2010) also considered the problem addressed in this work, but they proposed estimators of the total and the mean derived 
under the assumption of homogeneous link probabilities. Thus, our work is an extension of theirs. 

The structure of this paper is as follows. In Section 2 we introduce the LTS variant proposed by F\'elix-Medina and Thompson (2004), as
well as the notation to be used throughout this paper. In Section 3 we present the models and the MLEs of the population sizes proposed 
by F\'elix-Medina et al. (2015). In Section 4 we develop the strategy to construct the proposed model-based HTLEs and HKLEs of the total 
and the mean. In Section 5 we describe the variant of bootstrap that is proposed to construct the estimators of the variances of the
HTLEs and HKLEs of the totals and means, as well as the confidence intervals of these parameters. In Section 6 we present the results of 
two numerical studies carried out to observe the performance of the proposed estimators and confidence intervals and to compare their 
performance with that of the proposed by F\'elix-Medina and Monjardin (2010). Finally, in Section 7 we state some conclusions and 
suggestions for future research.

\section{Sampling design and notation\label{notation}} 

In this work we consider the variant of LTS proposed by F\'elix-Medina and Thompson (2004) which we will describe next. Let $U$ be a finite 
population of an unknown number $\tau$ of people. A portion $U_1$ of $U$ is assumed to be covered by a sampling frame of $N$ venues $A_1,
\ldots,A_N$, where the members of the population can be found with high probability. As in ordinary cluster sampling, each person in $U_1$ 
is assumed that can be assigned to only one venue in the frame, for instance, the venue where he or she spends most of his or her time. Notice 
that this does not imply that a person could not be found in different venues. Let $m_i$ denote the number of members of the population that 
belong, that is, that are assigned to the venue $A_i$, $i=1,\ldots,N$. From the previous assumption it follows that the number of people in 
$U_1$ is $\tau_1=\sum_1^Nm_i$ and the number of people in the portion $U_2=U-U_1$ of $U$ that is not covered by the frame is $\tau_2=\tau-
\tau_1$.

The sampling design is as follows. A SRSWOR $S_A$ of $n$ venues $A_1,\ldots,A_n$ is selected from the frame. The $m_i$ members of the population 
who belong to the sampled venue $A_i$ are identified and their associated $y$-values of a variable of interest $y$ are recorded, $i=1,\ldots,n$. 
Let $S_0$ be the set of people in the initial sample. Notice that the size of $S_0$ is $m=\sum_1^nm_i$. The people in each sampled venue are asked 
to name other members of the population. We will say that a person and a venue are linked if any of the people who belong to that venue names him 
or her. Let $x_{ij}^{(k)}=1$ if person $j\in U_k-A_i$ is linked to venue $A_i\in S_A$ and $x_{ij}^{(k)}=0$ if $j\in A_i$ or $j$ is not linked to 
$A_i$, $i=1,\ldots,n$; $k=1,2$. For each named person, the following information is recorded: the value of the variable of interest $y$ associated 
with him or her, the sampled venues that are linked to him or her, and the subset of $U$: $U_1-S_0$, a specific $A_i\in S_A$ or $U_2$, that 
contains him or her. Let $S_1$ be the set of people in $U_1-S_0$ who are linked to at least one venue in $S_A$, and let $S_2$ be the set of people 
in $U_2$ who are linked to at least one venue in $S_A$. We will denote by $r_k$ the size of $S_k$, $k=1,2$. Finally, let $S_1^{*}=S_0\cup S_1$ 
and $S_2^{*}=S_2$ be the sets of the sampled people from $U_1$ and $U_2$, respectively. Notice that the respective sizes of these sets are $m+r_1$ 
and $r_2$.

We will end this section by making the following remark. Spreen and Bogaerts (2015) recently proposed a 
sampling procedure to estimate the size of a hidden population called B-Graph sampling, which could be 
considered as a particular case of the LTS variant proposed by F\'elix-Medina and Thompson (2004). To 
see this, let us firstly describe the B-Graph sampling procedure. In this method a set of incomplete 
sampling frames of the population of interest is obtained. For instance, in the case of a population of 
drug users, two incomplete sampling frames could be the register of drug users who were detained by the 
police and that of drug users who were attended by health clinics. An initial SRSWOR of elements is 
selected from the pooled sampling frame, which is the one formed by the union of the several sampling 
frames. Then, every sampled element is asked to name his or her contacts who are also members of the 
population, either they are or not in the pooled frame. For every named person, the number of elements 
in the initial sample who name him or her is obtained. To estimate the size of the portion of the 
population that is not included in the pooled frame, those authors made the following analogy between 
B-Graph sampling and multiple capture-recapture sampling: each person in the initial sample of their
sampling design is considered as a sampling occasion of the capture-recapture procedure. Thus, the size
of the portion of the population that is not covered by the pooled frame is estimated by using a 
capture-recapture estimator, for instance, any of the proposed by Chao (1987), Chao (1988) or Zelterman 
(1988). The size of the entire population is estimated by the sum of the number of elements in the pooled 
frame and the estimate obtained by the capture-recapture estimator. Thus, if in the variant of LTS proposed 
by F\'elix-Medina and Thompson (2004), $\tau_1$ were known, and every $m_i$ were equal to 1, so that 
$N=\tau_1$, then the only problem would be to estimate $\tau_2$, which is the problem considered by Spreen 
and Bogaerts (2015). The difference is that those authors used Chao's (1988) estimator, whereas F\'elix-Medina 
and Thompson (2004) and F\'elix-Medina et al. (2015) used MLEs.

\section{Maximum likelihood estimators of the population sizes\label{MLES}} 

F\'elix-Medina et al. (2015) proposed MLEs of the population sizes $\tau_1$, $\tau_2$ and
$\tau$, which were derived from the following assumptions. The values $m_1,\ldots, m_N$ are
considered as realizations of the random variables $M_1,\ldots, M_N$, which are supposed to be 
independent and identically distributed Poisson random variables with mean $\lambda_1$. This 
implies that the joint conditional distribution of the vector of variables 
$\mathbf M_s=(M_1,\ldots,M_n,\tau_1-M)$, where $M=\sum_1^nM_i$, given that $\sum_1^NM_i=\tau_1$, 
is multinomial with parameter of size $\tau_1$ and vector of probabilities $(1/N,\ldots,1/N,1-n/N)$. 
The values $x_{ij}^{(k)}$s are assumed to be realizations of the random variables $X_{ij}^{(k)}$s, 
which given the sample $S_A$ of venues, are supposed to be independent Bernoulli random variables 
with means $p_{ij}^{(k)}$s, where the means or link probabilities $p_{ij}^{(k)}$s are given by the 
following Rasch model: 
{\setlength\arraycolsep{1pt}
\begin{equation}
\label{rasch}p_{ij}^{(k)}=\Pr(X^{(k)}_{ij}=1|S_A,\alpha_{(k)i},\beta_{(k)j})=\frac{\exp(\alpha_{(k)i}+
	\beta_{(k)j})}{1+\exp(\alpha_{(k)i}+\beta_{(k)j})},\, j\in U_k-A_i;\quad i=1,\ldots,n.
\end{equation}}

As is indicated in F\'elix-Medina et al. (2015), this model was considered by Coull and 
Agresti (1999) in the context of multiple capture-recapture sampling. The parameter
$\alpha_{(k)i}$ is a fixed (not random) effect that represents the potential that the 
venue $A_i$ has of forming links with people in $U_k-A_i$, and $\beta_{(k)j}$ is a 
random effect that represents the propensity of the person $j\in U_k$ to be linked to a 
sampled venue. Those authors suppose that $\beta_{(k)j}$ is normally distributed with mean $0$ 
and unknown variance $\sigma_k^2$ and that these variables are independent. The parameter 
$\sigma_k^2$ determines the degree of heterogeneity of the $p_{ij}^{(k)}$'s: great values 
of $\sigma_k^2$ imply high degrees of heterogeneity.

Henceforth all probability statements will be conditioned on the sample $S_A$ of venues
unless otherwise is specified. Now, let ${\mathbf X}_j^{(k)}=(X_{1j}^{(k)},\ldots,X_{nj}^{(k)})$ 
be the $n$-dimensional vector of link indicator variables $X_{ij}^{(k)}$s associated with the 
$j$-th person in $U_k-S_0$, and let $\Omega=\{\mathbf x=(x_1,\ldots,x_n)\in\mathbb{R}^n: x_i=0
\textrm{ or } x_i=1; i=1,\ldots,n\}$. Then the probability that ${\mathbf X}_j^{(k)}$ equals 
$\mathbf x=(x_1,\ldots,x_n)\in\Omega$, that is, the probability that the $j$-th person in $U_k-S_0$ is 
linked to only the venues $A_i\in S_A$ such that the $i$-th element $x_i$ of $\mathbf x$ equals 1, 
is
$$
\Pr({\mathbf X}_j^{(k)}={\mathbf x}|\boldsymbol\alpha_k,\beta_{(k)j})=\prod_{i=1}^n[p_{ij}^{(k)}]^{x_i}
[1-p_{ij}^{(k)}]^{1-x_i}=\prod_{i=1}^n\frac{\exp[x_i(\alpha_{(k)i}+\beta_{(k)j})]}
{1+\exp(\alpha_{(k)i}+\beta_{(k)j})},
$$
where $\boldsymbol\alpha_k=(\alpha_{(k)1},\ldots,\alpha_{(k)n})$. Therefore, the probability that the 
vector of link indicator variables associated with a randomly selected person in $U_k-S_0$ equals 
${\mathbf x}$ is
$$
\pi_{(k)\mathbf x}(\boldsymbol\alpha_k,\sigma_k)=\int\prod_{i=1}^n
\frac{\exp[x_i(\alpha_{(k)i}+\sigma_kz)]}
{1+\exp(\alpha_{(k)i}+\sigma_kz)}\phi(z)dz,
$$
where $\phi(\cdot)$ denotes the probability density function of the standard normal distribution [$N(0,$ $1)$].

F\'elix-Medina et al. (2015), following Coull and Agresti (1999), use the Gaussian quadrature method to
obtain the following approximation to $\pi_{(k)\mathbf x}$:
\begin{equation}
\label{gauss1}\tilde\pi_{(k)\mathbf x}=\tilde\pi_{(k)\mathbf x}(\boldsymbol\alpha_k,\sigma_k)=\sum_{t=1}^q\prod_{i=1}^n
\frac{\exp[x_i(\alpha_{(k)i}+\sigma_k z_t)]}{1+\exp(\alpha_{(k)i}+\sigma_k z_t)}\nu_t,
\end{equation} 
where $q$ is a fixed constant and $\{z_{t}\}$ and $\{\nu_t\}$ are obtained from tables (see Table 25.5 in Abramowitz and Stegun 1964)
or statistical software (see R library statmod developed by Giner and Smyth 2016).

Similarly, for person $j$ in $A_{i}\in S_{A}$, let $\mathbf{X}_{j}^{(A_{i})}=(X_{1j}^{(A_{i})},\ldots,
$ $X_{i-1j}^{(A_{i})},X_{i+1j}^{(A_{i})},\ldots ,X_{nj}^{(A_{i})})$ be the $(n-1)$-dimensional 
vector of link indicator variables $X_{i^\prime j}^{(A_{i})}$s associated with that person, and let
$\Omega_{-i}=\{\mathbf x=(x_1,\ldots,x_{i-1},x_{i+1},\ldots,x_n)\in\mathbb{R}^{n-1}: x_j=0\textrm{ or } x_j=1;
j=1,\ldots,n; j\ne i\}$. Then, the probability that $\mathbf{X}_{j}^{(A_{i})}$ equals $\mathbf x=(x_1,\ldots,x_{i-1},
x_{i+1},\ldots,x_n)\in\Omega_{-i}$, that is, the probability that the $j$-th person in $A_i$ is linked to only 
the venues $A_{i^\prime}\in S_A$, $i^\prime\neq i$, such that the $i^\prime$-th element $x_{i^\prime}$ of 
$\mathbf x$ equals 1, is
$$
\Pr({\mathbf X}_j^{(A_{i})}={\mathbf x}|\boldsymbol\alpha_1,\beta_{(1)j})=\prod_{i^\prime\neq i}^n[p_{i^\prime j}^{(1)}]^{x_{i^\prime}}
[1-p_{i^\prime j}^{(1)}]^{1-x_{i^\prime}}=\prod_{i^\prime\neq i}^n\frac{\exp[x_{i^\prime}(\alpha_{(1)i^\prime}+\beta_{(1)j})]}
{1+\exp(\alpha_{(1)i^\prime}+\beta_{(1)j})}
$$
and the Gaussian quadrature approximation to the probability $\pi_{(A_i)\mathbf x}
(\boldsymbol\alpha_1,$ $\sigma_1)$ that the vector of link indicator variables associated 
with a randomly person selected from the sampled venue $A_i$ equals the $(n-1)$-dimensional 
vector $\mathbf x=(x_1,\ldots,x_{i-1},x_{i+1},\ldots,x_n)$ is
\begin{equation}
\label{gauss2}\tilde\pi_{(A_i)\mathbf x}=\tilde\pi_{(A_i)\mathbf x}(\boldsymbol\alpha_1,\sigma_1)=\sum_{t=1}^q\prod_
{i^\prime\ne i}^n\frac{\exp[x_{i^\prime}(\alpha_{(1)i^\prime}+\sigma_1 z_t)]}{1+\exp(\alpha_{(1)i^\prime}+\sigma_1 z_t)}\nu_t.
\end{equation}

Under the previous assumptions, F\'elix-Medina et al. (2015) construct the likelihood function of $\tau_k$, $\boldsymbol\alpha_k$ 
and $\sigma_k$, $k=1,2$, which is proportional to a product of several multinomial distributions. One multinomial distribution is the
conditional distribution of the vector of variables $\mathbf M_s=(M_1,\ldots,M_n,\tau_1-M)$, given that $\sum_1^NM_i=\tau_1$, and that 
was indicated at the beginning of this section. Two others are the multinomial distributions of the vector of variables 
$\left(R_{\mathbf x}^{(k)}\right)_{\mathbf x\in\Omega}$, $k=1,2$, where $R_{\mathbf x}^{(1)}$ and $R_{\mathbf x}^{(2)}$ denote the 
number of people in $U_1-S_0$ and in $U_2$, respectively, whose vectors of link indicator variables $X_{ij}^{(k)}$, $k=1,2$, are equal 
to the vector $\mathbf x\in\Omega$. These distributions have parameters of size $\tau_1-m$, in the case of $k=1$, and $\tau_2$, in the 
case of $k=2$, and vector of probabilities $\left(\tilde\pi_{(k)\mathbf x}\right)_{\mathbf x\in\Omega}$, $k=1,2$. Finally, for each 
$i=1,\ldots,n$, we have a multinomial distribution of the vector of variables $\left(R_{\mathbf x}^{(A_i)}\right)_{\mathbf x\in\Omega_{-i}}$, 
where $R_{\mathbf x}^{(A_i)}$ denotes the number of people in $A_i$ whose vectors of link indicator variables $X_{ij}^{(A_i)}$ are equal to 
the vector $\mathbf x\in\Omega_{-i}$. This distribution has parameters of size $m_i$ and vector of probabilities 
$\left(\tilde\pi_{(A_i)\mathbf x}\right)_{\mathbf x\in\Omega_{-i}}$. Those authors proposed maximum likelihoods estimators of 
$\tau_k$, $\boldsymbol\alpha_k$ and $\sigma_k$, $k=1,2$, whose values are obtained by numerically maximizing the likelihood function. They 
called these estimators unconditional maximum likelihood estimators (UMLE) and denoted them as $\hat\tau_k^{(U)}$, 
$\boldsymbol{\hat\alpha}_k^{(U)}$ and $\hat\sigma_k^{(U)}$, $k=1,2$. Although these estimators do not have closed forms, the authors 
provided the following asymptotic approximations for $\hat\tau_1^{(U)}$ and $\hat\tau_2^{(U)}$:
\begin{equation}
\label{MLEtaus}\hat\tau_1^{(U)}=\frac{M+R_1}{1-(1-n/N){\tilde\pi}_{(1)\mathbf 0}
	(\boldsymbol{\hat\alpha}_1^{(U)},\hat\sigma_1^{(U)})}
\quad\textrm{and}\quad
\hat\tau_2^{(U)}=\frac{R_2}{1-{\tilde\pi}_{(2)\mathbf 0}(\boldsymbol{\hat\alpha}_2^{(U)},
	\hat\sigma_2^{(U)})},
\end{equation}
where $R_1$ and $R_2$ denote the numbers of distinct people in $U_1-S_0$ and $U_2$, respectively, that are linked to at least one venue in
$S_A$. Notice that these are not close forms because $\boldsymbol{\hat\alpha}_k^{(U)}$ and $\hat\sigma_k^{(U)}$ depend on $\hat\tau_k^{(U)}$. 
Once $\hat\tau_1^{(U)}$ and $\hat\tau_2^{(U)}$ are obtained, the UMLE of $\tau$ is $\hat\tau^{(U)}=\hat\tau_1^{(U)}+\hat\tau_2^{(U)}$.

Also, F\'elix-Medina et al. (2015), following Coull and Agresti (1999), used Sanathanan's
(1972) approach to derive conditional MLEs $\boldsymbol{\hat\alpha}_k^{(C)}$ and $\hat\sigma_k^{(C)}$ 
of $\boldsymbol\alpha_k$ and $\sigma_k$, $k=1,2$, given $R_k$. The values of these estimators
are obtained by maximizing numerically the corresponding conditional likelihood functions.
Then, they showed that the conditional MLEs $\hat\tau_1^{(C)}$ and $\hat\tau_2^{(C)}$ of $\tau_1$ and 
$\tau_2$ are given by $(\ref{MLEtaus})$, but replacing $\boldsymbol{\hat\alpha}_k^{(U)}$ and 
$\hat\sigma_k^{(U)}$ by $\boldsymbol{\hat\alpha}_k^{(C)}$ and $\hat\sigma_k^{(C)}$. Notice that in this
case the expressions $(\ref{MLEtaus})$ are closed forms. The conditional MLE of $\tau$ is $\hat\tau^{(C)}
=\hat\tau_1^{(C)}+\hat\tau_2^{(C)}$.

\section{Estimators of the total and mean}

In this section we will focus on the problem of estimating the total and the mean of the 
values of a variable of interest $y$. Let $y_j^{(k)}$ be the value of $y$ associated with
the $j$-th element of $U_k$, $j=1,\ldots,\tau_k$, $k=1,2$. In this work we will suppose
that the $y$-values are fixed numbers and not random variables. Notice that this 
assumption is the one made in traditional sampling. Then $Y_k=\sum_{j\in U_k}y_j^{(k)}$
and $\bar Y_k=Y_k/\tau_k$ represent the total and the mean of the portion $U_k$, $k=1,2$,
of the population. Similarly, $Y=Y_1+Y_2$ and $\bar Y=Y/\tau$ represent the total and the 
mean of the whole population $U$.

We cannot compute the design-based inclusion probabilities of the sampled elements because 
we do not know the venues in the frame that are linked to each sampled person, we however can
compute conditional model-based inclusion probabilities given the venues $A_i\in S_A$. These
probabilities are given by
{\setlength\arraycolsep{1pt}
\begin{eqnarray}
\label{pij1}\pi_{(1)j}(\boldsymbol\alpha_1,\sigma_1,\beta_{(1)j})&=&1-(1-n/N)\prod_{i=1}^n
(1-p_{ij}^{(1)})\quad\textrm{if } j\in U_1,\quad\textrm{and}
\\
\label{pij2}\pi_{(2)j}(\boldsymbol\alpha_2,\sigma_2,\beta_{(2)j})&=&1-\prod_{i=1}^n(1-p_{ij}^{(2)})
\quad\textrm{if } j\in U_2.
\end{eqnarray}}

The probabilities $\pi_{(k)j}(\boldsymbol\alpha_k,\sigma_k,\beta_{(k)j})$s are not known because
depend on unknown parameters. However, we could estimate them by estimating those parameters 
and replacing in $(\ref{pij1})$ and $(\ref{pij2})$ the parameters by their estimates. Both UMLEs and CMLEs
of $\boldsymbol\alpha_k$ and $\sigma_k$ have already been derived by F\'elix-Medina et al. (2015). We
will next derive a predictor of the random effect $\beta_{(k)j}$.

Thus, if the element $j\in U_k-S_0$, $k=1,2$, then the conditional joint probability density function of the vector 
$\mathbf X_j^{(k)}$ of link indicator variables associated with that element and the random effect $\beta_{(k)j}$ is
{\setlength\arraycolsep{1pt}
\begin{eqnarray}
f(\mathbf x_j^{(k)},\beta_{(k)j}|\boldsymbol\alpha_k,\sigma_k)&=&\Pr(\mathbf X_j^{(k)}=\mathbf x_j^{(k)}|\beta_{(k)j},
\boldsymbol\alpha_k) f(\beta_{(k)j}|\sigma_k)
\nonumber \\ \nonumber \\
&\propto&\prod_{i=1}^n[p_{ij}^{(k)}]^{x_{ij}^{(k)}}[1-p_{ij}^{(k)}]^{1-x_{ij}^{(k)}}
\exp[-(\beta_{(k)j})^2/2\sigma_k^2],
\nonumber
\end{eqnarray}}
whereas if the element $j\in A_{i'}\in S_A,\, i'=1,\dots,n$, then
$$
f(\mathbf x_j^{(A_{i'})},\beta_{(1)j}|\boldsymbol\alpha_1,\sigma_1)\propto\prod_{i\ne i'}^n[p_{ij}^{(1)}]
^{x_{ij}^{(A_{i'})}}[1-p_{ij}^{(1)}]^{1-x_{ij}^{(A_{i'})}}
\exp[-(\beta_{(1)j})^2/2\sigma_1^2].
$$
We will propose as a predictor of $\beta_{(k)j}$ the conditional expected value of $\beta_{(k)j}$ given
$\mathbf X_j^{(k)}$ $=\mathbf x_j^{(k)}$, evaluated either at the UMLEs $\boldsymbol{\hat\alpha}_k^{(U)}$ 
and $\hat\sigma_k^{(U)}$ or at the CMLEs $\boldsymbol{\hat\alpha}_k^{(C)}$ and $\hat\sigma_k^{(C)}$, that is
$$
\hat{\bar\beta}_{(k)j}^{(a)}=E(\beta_{(k)j}|\mathbf x_j^{(k)},\boldsymbol{\hat\alpha}^{(a)}_k,\hat\sigma_k^{(a)})=
\frac{\int\beta_{(k)j}f(\mathbf x_j^{(k)},\beta_{(k)j}|\boldsymbol{\hat\alpha}^{(a)}_k,\hat\sigma_k^{(a)})d\beta_{(k)j}}
{\int f(\mathbf x_j^{(k)},\beta_{(k)j}|\boldsymbol{\hat\alpha}^{(a)}_k,\hat\sigma_k^{(a)})d\beta_{(k)j}}, \qquad a=U, C.
$$
We will approximate $\hat{\bar\beta}_{(k)j}^{(a)}$ by using the Gaussian quadrature method, that is by
{\setlength\arraycolsep{1pt}
\begin{eqnarray}
\tilde{\bar\beta}_{(k)j}^{(a)}&=&
\frac{\hat\sigma_k^{(a)}\sum_{t=1}^q z_t\prod_{i=1}^n
\left\{\exp[x_i(\hat\alpha^{(a)}_{(k)i}+\hat\sigma_k^{(a)} z_t)]/[1+\exp(\hat\alpha^{(a)}_{(k)i}+\hat\sigma_k^{(a)} z_t)]\right\}\nu_t}
{\sum_{t=1}^q\prod_{i=1}^n
\left\{\exp[x_i(\hat\alpha^{(a)}_{(k)i}+\hat\sigma_k^{(a)} z_t)]/[1+\exp(\hat\alpha^{(a)}_{(k)i}+\hat\sigma_k^{(a)} z_t)]\right\}\nu_t}
\nonumber \\ \nonumber \\
&=&\frac{\hat\sigma_k^{(a)}\sum_{t=1}^q z_t\left\{\exp(\hat\sigma_k^{(a)} z_t\sum_{i=1}^nx_i)/
\prod_{i=1}^n[1+\exp(\hat\alpha^{(a)}_{(k)i}+\hat\sigma_k^{(a)} z_t)]\right\}\nu_t}
{\sum_{t=1}^q\left\{\exp(\hat\sigma_k^{(a)} z_t\sum_{i=1}^nx_i)/
\prod_{i=1}^n[1+\exp(\hat\alpha^{(a)}_{(k)i}+\hat\sigma_k^{(a)} z_t)]\right\}\nu_t}, \, a=U, C,
\nonumber \\
\label{hat-beta}& &
\end{eqnarray}}
if $j\in U_k-S_0$, $k=1,2$, and by
$$
\tilde{\bar\beta}_{(1)j}^{(a)}=\frac{\hat\sigma_1^{(a)}\sum_{t=1}^q z_t\left\{\exp(\hat\sigma_1^{(a)} z_t\sum_{i\ne i'}^nx_i)/
\prod_{i\ne i'}^n[1+\exp(\hat\alpha^{(a)}_{(1)i}+\hat\sigma_1^{(a)} z_t)]\right\}\nu_t}
{\sum_{t=1}^q\left\{\exp(\hat\sigma_1^{(a)} z_t\sum_{i\ne i'}^nx_i)/
\prod_{i\ne i'}^n[1+\exp(\hat\alpha^{(a)}_{(1)i}+\hat\sigma_1^{(a)} z_t)]\right\}\nu_t}, \, a=U, C,
$$
if $j\in A_{i'}\in S_A$, $i'=1,\dots,n$.

The previous expressions imply that $\tilde{\bar\beta}_{(k)j}^{(a)}$ depends on the $x_i$s through their sum, that is, on the
number of venues that are linked to the element $j$, but not on the particular venues to which that element is linked. Thus, 
if two persons $j$ and $j'$ in $U_k-S_0$ are linked to the same number of venues in $S_A$, the predictors 
$\tilde{\bar\beta}_{(k)j}$ and $\tilde{\bar\beta}_{(k)j'}$ are equal one another. The same happens for two persons in 
$A_i\in S_A$.

Thus, model-based Horvitz-Thompson-like estimators (HTLEs) of the totals $Y_k$, $k=1,2$, and $Y$ are
{\setlength\arraycolsep{1pt}
\begin{eqnarray}
\nonumber\hat Y_{HT.k}^{(a)}&=&\sum_{j\in S_k^*}y_j^{(k)}/\hat\pi_{(k)j}^{(a)}(\boldsymbol{\hat\alpha}^{(a)}_k,
\hat\sigma_k^{(a)},\tilde{\bar\beta}_{(k)j}^{(a)}),\, k=1,2,\,\textrm{and}\, \hat{Y}_{HT}^{(a)}=
\hat Y_{HT.1}^{(a)} + \hat Y_{HT.2}^{(a)}, \, a=U, C. \\
\label{YHT}&&
\end{eqnarray}}
Similarly, model-based HTLEs of the means $\bar Y_k$ and $\bar Y$ are
$$
\hat{\bar Y}_{HT.k}^{(a)}=\hat Y_{HT.k}^{(a)}/\hat\tau_k^{(a)},\, k=1,2,\quad\textrm{and}\quad 
\hat{\bar Y}_{HT}^{(a)}=\hat Y_{HT}^{(a)}/\hat\tau^{(a)}, \quad a=U, C.
$$
Notice that if we set $y_j^{(k)}=1$, for $j=1,\ldots,\tau_k$ and $k=1,2$, then $Y_k=\tau_k$, $k=1,2$, 
and $Y=\tau$. Therefore, HTLEs of $\tau_k$ and $\tau$ are $\hat\tau_{HT.k}^{(a)}=\hat Y_{HT.k}^{(a)}$,
$k=1,2$, and $\hat\tau_{HT}^{(a)}=\hat Y_{HT}^{(a)}$, $a=U, C$, where $\hat Y_{HT.k}^{(a)}$ and $\hat Y_{HT}^{(a)}$ 
are given by $(\ref{YHT})$ with $y_j^{(k)}=1$.

We could also define H\'ajek-like estimators (HKLEs) of the population totals and means. Thus, HKLEs of the means 
$\bar Y_k$ and $\bar Y$ are
$$
\hat{\bar Y}_{HK.k}^{(a)}=\hat Y_{HT.k}^{(a)}/\hat\tau_{HT.k}^{(a)},\, k=1,2,\quad\textrm{and}\quad 
\hat{\bar Y}_{HK}^{(a)}=\hat Y_{HT}^{(a)}/\hat\tau_{HT}^{(a)}, \quad a=U, C,
$$
and HKLEs of the totals $Y_k$ and $Y$ are
$$
\hat Y_{HK.k}^{(a)}=\hat{\bar Y}_{HK.k}^{(a)}\hat\tau_k^{(a)},\,k=1,2,\quad\textrm{and}\quad
\hat Y_{HK}^{(a)}=\hat{\bar Y}_{HK}^{(a)}\hat\tau^{(a)}, \quad a=U, C.
$$
 
\section{Bootstrap variance estimators and confidence intervals}

We propose the use of bootstrap to construct estimators of the variances of the proposed estimators of
the totals and means, as well as confidence intervals (CIs) for those population parameters. The proposed
bootstrap variant is obtained by combining the bootstrap version for finite populations proposed by Booth 
et al. (1994) and the parametric bootstrap variant (see Davison and Hinkley, 1997, Ch. 2). This version of 
bootstrap is an extension of the one used by F\'elix-Medina et al. (2015) to construct CIs based on their
proposed MLEs of the population sizes.

Hereinafter, we will denote by $\lfloor x\rfloor$, the greatest integer less than or equal to 
$x\in\mathbb{R}$. The steps of the proposed bootstrap procedure are the following. (i) Construct a population 
vector $\mathbf m_{Boot}$ of $N$ values of $m_i$s by means of the following procedure. If $N/n$ is an integer,
repeat $N/n$ times the observed sample of $n$ cluster sizes $\mathbf m_s=\{m_1,\ldots,m_n\}$. If $N/n$ is 
not an integer, that is, if $N=an+b$, where $a$ and $b$, $b<n$, are positive integers, then repeat $a$ times 
$\mathbf m_s$ and add to this set a SRSWOR of $b$ values of $m_i$s selected from $\mathbf m_s$. If the sum
of the elements of the vector $\mathbf m_{Boot}$ is greater than the value $\hat\tau_1^{(a)}$, $a=U, C$,
(depending of the type of estimator that is being considered), delete one element at a time from 
$\mathbf m_{Boot}$ starting from the $N$-th element until the sum is less than or equal to $\hat\tau_1^{(a)}$. 
Let $N_{Boot}$ be the final number of elements in $\mathbf m_{Boot}$. (ii) For each $k=1,2$, construct 
a population vector $\boldsymbol{\hat\alpha}_{(k)Boot}^{(a)}$ of length $N_{Boot}$ whose elements are the estimates 
$\hat\alpha_{(k)i}^{(a)}$s of the $\alpha_{(k)i}$s associated with the clusters whose sizes $m_i$s are in 
$\mathbf m_{Boot}$. (iii) For each $k=1,2$, construct a population vector $\boldsymbol{\hat\beta}_{(k)Boot}^{(a)}$ 
of length $\hat\tau_k^{(a)}$ whose first $m+r_1$ elements in the case of $k=1$, or whose first $r_2$ elements 
in the case of $k=2$, are the estimates $\hat{\bar\beta}_{(k)j}^{(a)}$s of the $\beta_{(k)j}$s associated with 
the people in $S_k^{*}$, and each one of the remaining elements is the estimate $\hat{\bar\beta}_{(k)0}^{(a)}$ 
of $\beta_{(k)j}$ obtained using $(\ref{hat-beta})$ with $x_i=0$, $i=1,\ldots,n$. (iv) For each $k=1,2$, 
construct a population vector $\mathbf{\hat y}_{(k)Boot}^{(a)}$ of length $\hat\tau_k^{(a)}$ whose first 
$m+r_1$ elements in the case of $k=1$, or whose first $r_2$ elements in the case of $k=2$, are the $y$-values 
associated with the elements in $S_k^{*}$, and the remaining elements are estimates of the $y$-values 
associated with the people in $U_k-S_k^{*}$ and obtained using the following procedure. If the variable of 
interest $y$ is continuous, then fit a simple linear regression model to the data $\left(\hat\pi_{(k)j}^{(a)}(\boldsymbol{\hat\alpha}^{(a)}_k,\hat\sigma_k^{(a)},\tilde{\bar\beta}_{(k)j}^{(a)}),
y_j^{(k)}\right)$, $j\in S_k^{*}$. Next, predict the $y$-value associated with the $j$-th element in $U_k-S_k^{*}$ 
by using a value sampled from the normal distribution with mean equals to the quantity obtained by evaluating the fitted
model at the estimate $\hat\pi_{(k)0}^{(a)}(\boldsymbol{\hat\alpha}^{(a)}_k,\hat\sigma_k^{(a)},\tilde{\bar\beta}_{(k)0}^{(a)})$
of the inclusion probability of an element in $U_k-S_k^{*}$ and variance equals to the estimate of the variance
of the error terms of the regression model. If the design matrix is numerically singular, then predict the $y$-value 
associated with $j\in U_k-S_k^{*}$ by a value sampled from the normal distribution with mean and variance given by 
the sample mean and sample variance, respectively, of the $y$-values associated with the elements in $S_k^{*}$. On 
the other hand, if the variable of interest $y$ is binary, then fit a simple logistic regression model to the data $\left(\hat\pi_{(k)j}^{(a)}(\boldsymbol{\hat\alpha}^{(a)}_k,\hat\sigma_k^{(a)},\tilde{\bar\beta}_{(k)j}^{(a)}),
y_j^{(k)}\right)$, $j\in S_k^{*}$. Next, predict the $y$-value associated with the $j$-th element in $U_k-S_k^{*}$ 
by using a value sampled from the Bernoulli distribution with success probability equals to the quantity obtained by 
evaluating the fitted model at $\hat\pi_{(k)0}^{(a)}(\boldsymbol{\hat\alpha}^{(a)}_k,\hat\sigma_k^{(a)},
\tilde{\bar\beta}_{(k)0}^{(a)})$. If the design matrix is numerically singular, then predict the $y$-value associated 
with $j\in U_k-S_k^{*}$ by a value sampled from the Bernoulli distribution with success probability equals to the sample 
mean of the $y$-values of the elements in $S_k^{*}$. (v) Select a SRSWOR of $n$ values $m_i$ from $\mathbf m_{Boot}$. 
Let $S_A^{Boot}=\{i_1,\dots,i_n\}$ be the set of indices of the $m_i$s in the sample. In addition, let 
$A_i^{Boot}=(\sum_{t=1}^{i-1}m_t,\sum_{t=1}^{i}m_t]\cap\mathbb Z$ be the set of indices $j$ associated with the elements 
in the cluster whose index is $i\in S_A^{Boot}$, where $m_t$ is the $t$-th element of $\mathbf m_{Boot}$ and $\mathbb Z$ 
is the set of the integer numbers. Finally, let $S_0^{Boot}=\cup_{i\in S_A^{Boot}}A_i^{Boot}$ be the set of indices $j$
associated with the elements in the clusters whose indices are in $ S_A^{Boot}$. (vi) For each $k=1,2$, $i\in S_A^{Boot}$ 
and $j\in \{1,\ldots,\lfloor\hat\tau^{(a)}_1\rfloor\}-A_i^{Boot}$ in the case of $k=1$, or 
$j\in\{1,\ldots,\lfloor\hat\tau_2^{(a)}\rfloor\}$ in the case of $k=2$, generate a value $x_{ij}^{(k)}$ by sampling from the 
Bernoulli distribution with success probability equals to the value obtained by evaluating $(\ref{rasch})$ at the $i$-th 
element of the vector $\boldsymbol{\hat\alpha}_{(k)Boot}^{(a)}$ and the $j$-th element of the vector 
$\boldsymbol{\hat\beta}_{(k)Boot}^{(a)}$. (vii) Compute the estimates of the sizes $\tau_1$, $\tau_2$ and $\tau$; those of
the totals $Y_1$, $Y_2$ and $Y$, and those of the means $\bar Y_1$, $\bar Y_2$ and $\bar Y$ using the same procedure as that 
used to compute the original estimates. (viii) Repeat the steps (v)-(vii) a large enough number $B$ of times.

To construct the CIs for the population totals and means we could use any of the different bootstrap alternatives that
have been proposed. For instance, if we did not want to assume any probability distribution for an estimator, we could
use the basic or the percentile method. (See Davison and Hinkley, 1997, Ch. 5, for descriptions of these methods.)
Although this type of alternative has good properties of robustness, it requires a large number $B$ of bootstrap samples, 
say $B=1000$, and this might be a serious problem if the estimator require much time to be computed. On the other
hand, if we were willing to assume a distribution probability for an estimator, we could use the $B$ bootstrap estimates
that were computed using that estimator to estimate its variance and construct the CI using the assumed distribution and the 
estimated variance. In this case, the number $B$ of required bootstrap samples is not so large, say $50\le B\le 200$ is 
generally enough. We will follow this approach using some ideas taken from F\'elix-Medina et al. (2015). 

Thus, as in that paper, we will estimate the variance of an estimator $\hat\theta$ of the population parameter $\theta$, by 
using Huber's proposal 2 to jointly estimate the parameters of location and scale from the bootstrap sample of $B$ values 
$\hat\theta_b$. (See Staudte and Sheather, 1990, Sec. 4.5, for a description of this method.) In particular, the estimate of 
the parameter of scale is an estimate of the standard deviation $\sqrt{\hat V(\hat\theta)}$ of $\hat\theta$. The idea behind 
the use of this estimator is that it yields an estimate of the standard deviation that is robust to very large values 
$\hat\theta_b$ which are likely to occur. 

To construct the CIs we will use the following approach. (i) If the parameter is $\tau_k$, $k=1,2$, or $\tau$, then, as in 
F\'elix-Medina et al. (2015), we will assume that $\hat\tau^{(a)}_k-\nu_k$ is lognormally distributed, where $\hat\tau_k^{(a)}$,
$a=U,C$, is an estimator of $\tau_k$ and $\nu_k$ is the number of sampled elements from $U_k$. Thus, a CI for $\tau_k$ is 
$\left(\nu_k+(\hat\tau_k^{(a)}-\nu_k)/c_k,\nu_k+(\hat\tau_k^{(a)}-\nu_k)\times c_k\right)$, where 
$c_k=\exp\left\{z_{\alpha/2}\sqrt{\ln[1+\hat V(\hat\tau^{(a)}_k)/(\hat\tau^{(a)}_k-\nu_k)^2]}\right\}$, $z_{\alpha/2}$ is the upper $\alpha/2$ 
point of the standard normal distribution and $\hat V(\hat\tau^{(a)}_k)$ is an estimate of the variance of $\hat\tau^{(a)}_k$. 
(See Williams et al., 2002, Sec. 14.2, for a description of this type of CI.) A CI for $\tau$ is built analogously. The values of 
$\nu_1$, $\nu_2$ and $\nu$ that are used in the CIs for $\tau_1$, $\tau_2$ and $\tau$ are $m+r_1$, $r_2$ and $m+r_1+r_2$, respectively. 
(ii) If the parameter is $\bar Y_k$, $k=1,2$, or $\bar Y$, and it is a proportion, that is, the $y$-value associated with an element is 
equal to one if the element has a characteristic of interest and is equal to zero otherwise, then we will assume that the number of sampled 
elements with the characteristic of interest has a binomial distribution and a CI for $\bar Y$ will be constructed using the proposal of 
Korn and Graubard (1998), which is an adaptation of the Clopper-Pearson CI for a proportion in the case of complex samples. Thus, a CI for 
$\bar Y_k$ is $\left(\nu_{(k)1}^{(a)}F_{\nu_{(k)1}^{(a)},\,\nu_{(k)2}^{(a)}}(\alpha /2)
[\nu_{(k)2}^{(a)}+\nu_{(k)1}^{(a)}F_{\nu_{(k)1}^{(a)},\,\nu_{(k)2}^{(a)}}(\alpha /2)],\nu_{(k)3}^{(a)}F_{\nu_{(k)3}^{(a)},\,
\nu_{(k)4}^{(a)}}(1-\alpha /2)[\nu_{(k)4}^{(a)}+\right.$ $\left. \nu_{(k)3}^{(a)}F_{\nu_{(k)3}^{(a)},\,\nu_{(k)4}^{(a)}}(1-\alpha /2)]\right)$, 
where $\nu_{(k)1}^{(a)}=2y_k^{(a)}$, $\nu_{(k)2}^{(a)}=2(n_k^{(a)}-y_k^{(a)}+1)$, $\nu_{(k)3}^{(a)}=2(y_k^{(a)}+1)$, 
$\nu_{(k)4}^{(a)}=2(n_k^{(a)}-y_k^{(a)})$, $y_k^{(a)}=n_k^{(a)}\hat{\bar Y}_k^{(a)}$, $n_k^{(a)}=\hat{\bar Y}_k^{(a)}
(1-\hat{\bar Y}_k^{(a)})/\hat V(\hat{\bar Y}_k^{(a)})$, $\hat{\bar Y}_k^{(a)}$ is an estimator of $\bar Y_k$, $\hat V(\hat{\bar Y}_k^{(a)})$ is 
an estimate of the variance of $\hat{\bar Y}_k^{(a)}$ and $F_{d_1,d_2}(\beta)$ is the $\beta$ quantile of the $F$ distribution with $d_1$ and 
$d_2$ degrees of freedom. A CI for $\bar Y$ is built analogously. (iii) If the parameter is $\bar Y_k$, $k=1,2$, or $\bar Y$, and it is the mean 
of the $y$-values a continuous variable of interest or if it is $Y_k$, $k=1,2$, or $Y$, that is, it is the total of the $y$-values of a continuous 
or a binary variable of interest, we will assume that the estimator $\hat{\bar Y}_k^{(a)}$ (or $\hat Y^{(a)}_k$) is normally distributed. Thus a CI 
for $\bar Y_k$ is $\left(\hat{\bar Y}_k^{(a)}-z_{\alpha/2}\sqrt{\hat V(\hat{\bar Y}_k^{(a)})},\hat{\bar Y}_k^{(a)}+z_{\alpha/2}
\sqrt{\hat V(\hat{\bar Y}_k^{(a)})}\right)$, where $z_{\alpha/2}$ is the upper $\alpha/2$ point of the standard normal distribution and 
$\hat V(\hat{\bar Y}_k^{(a)})$ is an estimate of the variance of $\hat{\bar Y}_k^{(a)}$. CIs for $Y_k$, $\bar Y$ and $Y$ are built analogously. 
	
\section{Monte Carlo studies}

In order to observe the performance of the proposed estimators and CIs and to compare their performance with the ones proposed by
F\'elix-Medina and Monjardin (2010), which were derived under the assumption of homogeneity of the link probabilities, we carried out
two numerical studies. In the first study we used data from the National Longitudinal Study of Adolescent Health collected during the 
1994-1995 school year to construct a population, whereas in the second one we used artificial data to construct two populations with 
specific characteristics. Both studies were carried out using the R software environment for statistical computing (R Core Team, 2018).

\subsection{Populations constructed using data from the National Longitudinal Study of Adolescent Health}

In this Monte Carlo Study we used data from the National Longitudinal Study of Adolescent Health (Add Health) to construct a population. The Add Health
is a longitudinal study of a representative sample of more than 90000 adolescents who in the years 1994-95 were in grades 7-12 in the United States. The 
participants were followed through adolescence and the transition to adulthood with the goal of helping to explain the causes of adolescent health and
health behavior. The sample of students was selected by a stratified probability proportional to size cluster sampling design, where the clusters were
the high schools and the strata were defined in terms of region, urbanicity, school type, etc. For each of the 84 selected high schools, one of its feeder
middle school was selected with probability proportional to the number of contributed students to the high school. Each student in the representative 
sample was asked to named up to 5 male and 5 female friends within his or her high school or in the feeder school, and in addition, to complete an 
in-school questionnaire. Thus, the collected information can be modeled as a directed network, where the nodes are the sampled students and their referred 
friends, and a directed arc from node $i$ to node $j$ is considered to exist if student $i$ names student $j$ as a friend. See Harris (2013) for a 
description of this study.

A subset of the data obtained in the Add Health study is contained in Linton Freeman's web page: http://moreno.ss.uci.edu/data.html\#ahealth. I our numerical
study we used data from this subset corresponding to the high school and its feeder in Community 50 to construct a population $U$ of $\tau=2497$ elements
divided into subpopulations $U_1$ and $U_2$ of sizes $\tau_1=1800$ and $\tau_2=697$, respectively. The elements assigned to $U_1$ were those at positions
labeled with odd numbers in the data file and that named at least one friend plus a simple random sample of elements at positions labeled with even numbers
and that named at least one friend. These elements were grouped into $N=150$ clusters of sizes $m_i$, $i=1,\ldots,N$, obtained by sampling from a negative 
binomial distribution with mean and variance equal to 12 and 24, respectively. The elements assigned to $U_2$ were the remaining elements in the data file 
that were named as a friend by at least one element in $U_1$. Once the subpopulation $U_k$ was constructed, the $N\times\tau_k$ matrix $\mathbf X_k$ of 
values $x_{ij}^{(k)}$s of the link indicator variables $X_{ij}^{(k)}$s was constructed, $k=1,2$. We considered as response variables the following: 
``Number of friends'' (named by each element) and ``Gender'' (1=male, 0=female). The totals and means of the variable ``Number of friends'' were 
$(Y_1,Y_2,Y)=(10101,2729,12830)$ and $(\bar Y_1,\bar Y_2,\bar Y)=(5.612,3.915,5.138)$, and those of the variable ``Gender'' were $(Y_1,Y_2,Y)=(838,361,1200)$ 
and $(\bar Y_1,\bar Y_2,\bar Y)=(0.466,0.518,0.481)$. For an initial sample size $n=20$, which was the size used in this study, the values of the Pearson 
correlation coefficients between the values of the variable ``Number of friends'' and those of the inclusion probabilities associated with subpopulations 
$U_1$ and $U_2$ were $\rho(y^{(1)},\pi_{(1)})\approx 0.36$ and $\rho(y^{(2)},\pi_{(2)})\approx 0.29$, respectively, whereas the corresponding values for 
the variable ``Gender'' were $\rho(y^{(1)},\pi_{(1)})\approx -0.04$ and $\rho(y^{(2)},\pi_{(2)})\approx -0.07$.

The Monte Carlo study was carried out by repeatedly selecting $r$ samples from the population $U$ using the sampling design described in
Section 2. Thus, a SRSWOR of $n=20$ values $m_i$ was selected from the population of $N=150$ values. For each selected value $m_i$, the
values $x_{ij}^{(k)}$s of the link indicator variables $X_{ij}^{(k)}$, $j=1,\ldots,\tau_k$, were obtained from the matrix $\mathbf X_k$, 
$k=1,2$. Furthermore, for each element $j\in U_k$ that was sampled, its associated values $y_j^{(k)}$s of both response variables were 
recorded. From each selected sample, the following estimators of the population sizes $\tau_1$, $\tau_2$ and $\tau$ were computed: the 
UMLEs and CMLEs $\hat\tau_1^{(a)}$, $\hat\tau_2^{(a)}$ and $\hat\tau^{(a)}$, $a=U, C$, proposed by F\'elix-Medina et al. (2015); both 
types of HTLEs $\hat\tau_{HT.1}^{(a)}$, $\hat\tau_{HT.2}^{(a)}$ and $\hat\tau_{HT}^{(a)}$, proposed in this work: the ones based on the 
UMLEs ($a=U$) and those based on the CMLEs ($a=C$) of the inclusion probabilities; the MLEs $\hat\tau_{ML.1}^{(H)}$, $\hat\tau_{ML.2}^{(H)}$ 
and $\hat\tau_{ML}^{(H)}$ proposed by F\'elix-Medina and Thompson (2004) and derived under the assumption of homogeneity of the link 
probabilities, as well as the Bayesian-assisted estimators $\hat\tau_{BA.1}^{(H)}$, $\hat\tau_{BA.2}^{(H)}$ and $\hat\tau_{BA}^{(H)}$ 
proposed by F\'elix-Medina and Monjardin (2006) and derived also under the homogeneity assumption. We also computed the following 
estimators of the population totals $Y_1$, $Y_2$ and $Y$, and means $\bar Y_1$, $\bar Y_2$ and $\bar Y$: the two types of HTLEs 
$\hat Y_{HT.1}^{(a)}$, $\hat Y_{HT.2}^{(a)}$ and $\hat Y_{HT}^{(a)}$ and $\hat{\bar Y}_{HT.1}^{(a)}$, $\hat{\bar Y}_{HT.2}^{(a)}$ and 
$\hat{\bar Y}_{HT}^{(a)}$, and the two types of HKLEs $\hat Y_{HK.1}^{(a)}$, $\hat Y_{HK.2}^{(a)}$ and $\hat Y_{HK}^{(a)}$ and 
$\hat{\bar Y}_{HK.1}^{(a)}$, $\hat{\bar Y}_{HK.2}^{(a)}$ and $\hat{\bar Y}_{HK}^{(a)}$ proposed in this work. One type was based on the 
UMLEs ($a=U$) and the other on the CMLEs ($a=C$) of the inclusion probabilities. Finally, the following two types of HTLEs of the population 
totals and means proposed by F\'elix-Medina and Monjardin (2010) and derived under the assumption of homogeneity of the link probabilities 
were computed: $\hat Y_{a.1}^{(H)}$, $\hat Y_{a.2}^{(H)}$ and $\hat Y_{a}^{(H)}$ and $\hat{\bar Y}_{a.1}^{(H)}$, $\hat{\bar Y}_{a.2}^{(H)}$ 
and $\hat{\bar Y}_{a}^{(H)}$. Here, one type was based on the MLEs ($a=ML$) and the other on the Bayesian-assisted estimators ($a=BA$) of of 
the inclusion probabilities.

The performance of an estimator $\hat\theta$ of a parameter $\theta$ was evaluated by its relative bias (r-bias), the square root of its
relative mean square error ($\sqrt{\textrm{r-mse}}$), the median of its relative estimation error (mdre), and the median of its absolute
relative estimation error (mdare) defined as $\textrm{r-bias}=\sum_1^r(\hat\theta_i-\theta)/(r\theta)$, $\sqrt{\textrm{r-mse}}=\sqrt{\sum_1^r
(\hat\theta_i-\theta)^2/(r\theta^2)}$, $\textrm{mdre}=\textrm{median}\{(\hat\theta_i-\theta)/\theta\}$ and $\textrm{mdare}=\textrm{median}
\{|(\hat\theta_i-\theta)/\theta|\}$, respectively, where $\hat\theta_i$, is the value of $\hat\theta$ obtained in the $i$-th sample,
$i=1,\ldots,r$. In the case of the point estimators of the population sizes, totals and means their performance was evaluated using $r=5000$
samples.

We also computed estimators of the variances of these point estimators. In the case of the estimators derived under the assumption of
heterogeneous link probabilities we used the bootstrap variance estimators described in Section 5 based on $B=50$ bootstrap samples. 
In the case of the estimators based on the assumption of homogeneous link probabilities we used the variance estimators described in the 
papers in which the estimators were presented. The performance of a variance estimator $\hat V(\hat\theta)$ of the variance $V(\hat\theta)$ 
of $\hat\theta$ was also evaluated by its r-bias, $\sqrt{\textrm{r-mse}}$, mdre and mdre, where $V(\hat\theta)$ was computed by the sample
variance of the $\hat\theta_i$, $i=1,\ldots,r$. Because of the time required to computed the bootstrap variance estimators we used $r=500$
samples, whereas in the case of the estimators of the variances of the point estimators derived under the assumption of homogeneous link 
probabilities we used $r=5000$ samples.

From each point estimator and its associated variance estimator a 95\% CI was computed for the corresponding parameter. In the case of the
estimators based on the assumption of heterogeneous link probabilities the CIs were computed as was described in Section 5, whereas in the 
case of those based on the homogeneity assumption the CIs were Wald type CIs. The performance of a CI was evaluated by its coverage 
probability (cp) defined as the proportion of samples in which the parameter is inside the interval, and by both its mean relative length 
(mrl) and median relative length (mdrl) defined as the sample mean and median of the lengths of the $r$ intervals divided by the value of 
the parameter, respectively. In the case of the CI based on point estimators derived under the assumption of heterogeneous link 
probabilities we used $r=500$ samples, whereas in the case of the estimators derived under the assumption of homogeneous link probabilities 
we used $r=5000$ samples.

It is worth noting that, in this and in the following study, in the case of the estimators that were derived under the assumption of 
heterogeneous link probabilities, we only present the outcomes corresponding to the estimators based on the UMLEs of these probabilities 
because their performance was very similar to that of the estimators based on the CMLEs of the link probabilities. In the case of 
the estimators derived under the assumption of homogeneous link probabilities, we only present the outcomes corresponding to the estimators 
based on the MLEs of these probabilities because their performance was very similar to that of the estimators based on the Bayesian assisted 
estimators of the link probabilities. In addition, in the descriptions of the results of numerical studies we will use the convention that 
the performance of a point estimator will be considered as acceptable if both its r-bias (or mdre) and its $\sqrt{\textrm{r-mse}}$ (or mdare) 
are around or are lesser than 0.1. Similarly, that the performance of a 95\% CI is acceptable if its cp is around or greater than 0.9 and its 
mrl (or mdrl) is around or is lesser than 0.4 ($=4\times 0.1$).

\renewcommand{\baselinestretch}{0.95} 
\small\normalsize

{\setlength\tabcolsep{1.5pt}
	\begin{table}
		\vspace{-.1in}
		\begin{center} 
			\caption{Simulation results obtained for the point and standard deviation 
				estimators in a population constructed using data from the National
				Longitudinal Study of Adolescent Health.}
			%			\begin{tabular*}{5.5in}{@{\extracolsep{0.27mm}}
			%					|cc|cccc@{\hspace{1pt}}|cccc@{\hspace{1pt}}|cccc@{\hspace{1pt}}|cccc@{\hspace{1pt}}|}
			\begin{tabular}{|cc|cccc|cccc||cccc|cccc|}
				\hline
				\multicolumn{2}{|c|}{} & \multicolumn{8}{c||}{Point estimators} & 
				\multicolumn{8}{c|}{Standard deviation estimators}  \\ 
				\hline
				\multicolumn{2}{|c|}{Resp. variable} & \multicolumn{4}{c|}{Num. of friends} & \multicolumn{4}{c||}{Gender} & 
				\multicolumn{4}{c|}{Num. of friends} & \multicolumn{4}{c|}{Gender} \\
				\hline			
				\multicolumn{2}{|c|} {\raisebox{3ex}{Estimator}} & \shortstack{\\r\\b\\i\\a\\s} & 
				$\sqrt{\textrm{\shortstack{r\\m\\s\\e}}}$ & \shortstack{m\\d\\r\\e} & \shortstack{m\\d\\a\\r\\e} &  
				\shortstack{\\r\\b\\i\\a\\s} & $\sqrt{\textrm{\shortstack{r\\m\\s\\e}}}$ & 
				\shortstack{m\\d\\r\\e} & \shortstack{m\\d\\a\\r\\e} &
				\shortstack{\\r\\b\\i\\a\\s} & $\sqrt{\textrm{\shortstack{r\\m\\s\\e}}}$ & 
				\shortstack{m\\d\\r\\e} & \shortstack{m\\d\\a\\r\\e} &
				\shortstack{\\r\\b\\i\\a\\s} & $\sqrt{\textrm{\shortstack{r\\m\\s\\e}}}$ & 
				\shortstack{m\\d\\r\\e} & \shortstack{m\\d\\a\\r\\e} \\
				\hline
				UMLEs & $\hat{\tau}_{1}^{(U)}$ & -$.01$ & $.06$ & -$.01$ & $.04$ & -$.01$ & $.06$ & -$.01$ & $.04$	
				& -$.15$ & $.23$ & -$.17$ & $.19$ & -$.15$ & $.23$ & -$.17$ & $.19$ \\ 
				of & $\hat{\tau}_{2}^{(U)}$ & $.06$ & $.25$ & $.00$ & $.13$ & $.06$ & $.25$ & $.00$ & $.13$
				& $.08$ & $1.3$ & -$.29$ & $.43$ & $.08$ & $1.3$ & -$.29$ & $.43$ \\ 
				sizes & $\hat{\tau}^{(U)}$ & $.01$ & $.08$ & $.00$ & $.05$ & $.01$ & $.08$ & $.00$ & $.05$
				& $.05$ & $.99$ & -$.24$ & $.33$ & $.05$ & $.99$ & -$.24$ & $.33$ \\ \hline
				HTLEs & $\hat{\tau}_{HT.1}^{(U)}$ & -$.04$ & $.07$ & -$.04$ & $.05$ & -$.04$ & $.07$ & -$.04$ & $.05$
				& -$.15$ & $.23$ & -$.16$ & $.18$ & -$.15$ & $.23$ & -$.16$ & $.18$ \\ 
				of & $\hat{\tau}_{HT.2}^{(U)}$ & -$.05$ & $.15$ & -$.07$ & $.10$ & -$.05$ & $.15$ & -$.07$ & $.10$
				& $.13$ & $.81$ & -$.10$ & $.31$ & $.13$ & $.81$ & -$.10$ & $.31$ \\ 
				sizes & $\hat{\tau}^{(U)}_{HT}$ & -$.05$ & $.08$ & -$.05$ & $.05$ & -$.05$ & $.08$ & -$.05$ & $.05$ 
				& -$.04$ & $.47$ & -$.17$ & $.25$ & -$.04$ & $.47$ & -$.17$ & $.25$ \\ \hline
				HTLEs & $\hat{Y}_{HT.1}^{(U)}$ & $.00$ & $.06$ & $.01$ & $.04$ & -$.07$ & $.09$ & -$.07$ & $.07$ 
				& -$.14$ & $.22$ & -$.16$ & $.17$ & -$.15$ & $.22$ & -$.16$ & $.17$  \\ 
				of & $\hat{Y}_{HT.2}^{(U)}$ & $.07$ & $.17$ & $.05$ & $.09$ & -$.06$ & $.16$ & -$.08$ & $.11$ 
				& $.12$ & $.75$ & -$.10$ & $.31$ & $.19$ & $.81$ & -$.04$ & $.29$ \\ 
				totals & $\hat{Y}_{HT}^{(U)}$ & $.02$ & $.06$ & $.02$ & $.04$ & -$.07$ & $.09$ & -$.07$ & $.07$ 
				& -$.07$ & $.36$ & -$.16$ & $.21$ & -$.02$ & $.45$ & -$.13$ & $.23$ \\ \hline
				HTLEs & $\hat{\bar Y}_{HT.1}^{(U)}$ & $.01$ & $.02$ & $.01$ & $.02$ & -$.06$ & $.07$ & -$.06$ & $.06$
				& -$.15$ & $.19$ & -$.15$ & $.16$ & -$.03$ & $.13$ & -$.03$ & $.09$  \\ 
				of & $\hat{\bar Y}_{HT.2}^{(U)}$ & $.03$ & $.10$ & $.04$ & $.08$ & -$.10$ & $.13$ & -$.09$ & $.09$
				& -$.14$ & .$42$ & -$.27$ & $.32$ & $.03$ & $.42$ & -$.10$ & $.24$ \\ 
				means & $\hat{\bar Y}_{HT}^{(U)}$ & $.01$ & $.08$ & $.02$ & $.03$ & -$.07$ & $.06$ & -$.07$ & $.07$
				& -$.09$ & $.70$ & -$.33$ & $.40$ & $.10$ & $.67$ & -$.09$ & $.20$ \\	\hline				
				
				HKLEs & $\hat{Y}_{HK.1}^{(U)}$ & $.04$ & $.08$ & $.04$ & $.05$ & -$.03$ & $.08$ & -$.03$ & $.05$ 
                & -$.15$ & $.23$ & -$.17$ & $.18$ & -$.16$ & $.23$ & -$.18$ & $.19$  \\ 
                of & $\hat{Y}_{HK.2}^{(U)}$ & $.19$ & $.33$ & $.13$ & $.15$ & $.04$ & $.24$ & -$.01$ & $.13$ 
                & $.05$ & $1.2$ & -$.30$ & $.43$ & $.12$ & $1.3$ & -$.26$ & $.40$ \\ 
                totals & $\hat{Y}_{HK}^{(U)}$ & $.08$ & $.11$ & $.07$ & $.07$ & -$.01$ & $.09$ & -$.02$ & $.06$ 
                & $.03$ & $.89$ & -$.22$ & $.30$ & $.06$ & $.96$ & -$.21$ & $.31$ \\ \hline
                HKLEs & $\hat{\bar Y}_{HK.1}^{(U)}$ & $.05$ & $.05$ & $.05$ & $.05$ & -$.02$ & $.04$ & -$.02$ & $.03$
                & -$.12$ & $.16$ & -$.13$ & $.13$ & -$.06$ & $.14$ & -$.06$ & $.09$  \\ 
                of & $\hat{\bar Y}_{HK.2}^{(U)}$ & $.13$ & $.13$ & $.13$ & $.13$ & -$.02$ & $.04$ & -$.02$ & $.03$
                & $.01$ & .$16$ & $.00$ & $.10$ & $.31$ & $.37$ & $.30$ & $.30$ \\ 
                means & $\hat{\bar Y}_{HK}^{(U)}$ & $.07$ & $.07$ & $.07$ & $.07$ & -$.02$ & $.03$ & -$.02$ & $.02$
                & -$.09$ & $.29$ & -$.15$ & $.20$ & $.01$ & $.16$ & -$.00$ & $.10$ \\	\hline
											
				Homo. & $\hat{\tau}_{1}^{(H)}$ & -$.14$ & $.15$ & -$.14$ & $.14$ & -$.14$ & $.15$ & -$.14$ & $.14$	
				& -$.42$ & $.43$ & -$.42$ & $.42$ & -$.42$ & $.43$ & -$.42$ & $.42$ \\ 
				MLEs & $\hat{\tau}_{2}^{(H)}$ & -$.24$ & $.25$ & -$.24$ & $.24$ & -$.24$ & $.25$ & -$.24$ & $.24$
				& -$.24$ & $.29$ & -$.26$ & $.26$ & -$.24$ & $.29$ & -$.26$ & $.26$ \\ 
				sizes & $\hat{\tau}^{(H)}$ & -$.17$ & $.18$ & -$.17$ & $.17$ & -$.17$ & $.18$ & -$.17$ & $.17$
				& -$.45$ & $.45$ & -$.45$ & $.45$  & -$.45$ & $.45$ & -$.45$ & $.45$ \\ \hline
				Homo. & $\hat{Y}_{HT.1}^{(H)}$ & -$.09$ & $.11$ & -$.09$ & $.09$ & -$.16$ & $.17$ & -$.16$ & $.16$ 
				& -$.20$ & $.21$ & -$.19$ & $.19$ & -$.22$ & $.23$ & -$.22$ & $.22$  \\ 
				HTLEs & $\hat{Y}_{HT.2}^{(H)}$ & -$.13$ & $.16$ & -$.14$ & $.14$ & -$.25$ & $.26$ & -$.25$ & $.25$ 
				& -$.17$ & $.23$ & -$.19$ & $.20$ & -$.19$ & $.25$ & -$.21$ & $.22$ \\ 
				totals & $\hat{Y}_{HT}^{(H)}$ & -$.10$ & $.11$ & -$.10$ & $.10$ & -$.19$ & $.20$ & -$.19$ & $.19$ 
				& -$.28$ & $.28$ & -$.27$ & $.27$ & -$.31$ & $.31$ & -$.31$ & $.31$ \\ \hline
				Homo. & $\hat{\bar Y}_{HT.1}^{(H)}$ & $.06$ & $.06$ & $.06$ & $.06$ & -$.02$ & $.04$ & -$.02$ & $.03$
				& -$.18$ & $.19$ & -$.18$ & $.18$ & -$.13$ & $.14$ & -$.13$ & $.13$  \\ 
				HTLEs & $\hat{\bar Y}_{HT.2}^{(H)}$ & $.14$ & $.14$ & $.14$ & $.14$ & -$.02$ & $.04$ & -$.02$ & $.03$
				& -$.10$ & .$12$ & -$.11$ & $.11$ & $.12$ & $.14$ & $.12$ & $.12$ \\ 
				means & $\hat{\bar Y}_{HT}^{(H)}$ & $.08$ & $.02$ & $.08$ & $.08$ & -$.02$ & $.03$ & -$.02$ & $.03$
				& -$.25$ & $.26$ & -$.26$ & $.26$ & -$.13$ & $.14$ & -$.13$ & $.13$ \\
				\hline				
				%\multicolumn{10}{l}{Notes: $\hat\tau_k$, and $\hat\tau$, MLEs derived under the 
				% heterogeneity assumption; $\tilde\tau_k$ and} \\
				\multicolumn{18}{l}{Notes: Results for point estimators are based on 5000 samples; those for sd estimators} \\
				\multicolumn{18}{l}{derived under the heterogeneity assumption are based on 500 samples and those derived} \\
				\multicolumn{18}{l}{under the homogeneity assumption are based on 5000 samples. Average sampling frac-} \\
				\multicolumn{18}{l}{tions are $f_1=0.46$ and $f_2=0.40$. No convergence problems were observed, except in one} \\
				\multicolumn{18}{l}{sample in which the point estimators of the parameters of $U_2$ could not be computed.} \\ 
				%			\end{tabular*}
			\end{tabular}		
		\end{center}
\end{table}}

\renewcommand{\baselinestretch}{1.0} 
\small\normalsize

The results of the study on the point estimators are shown in Table 1 and in Figure 1. We can see that regardless of the type of response variable, 
the performance of each estimator that was derived under the heterogeneity assumption was pretty acceptable. The values of its mdre and mdare 
(summary statistics that are not affected by large values of the estimators) were generally small. The exceptions were the estimators 
$\hat{Y}_{HK.2}^{(U)}$ and $\hat{\bar Y}_{HK.2}^{(U)}$, which in the case of the response variable ``Number of friends'' presented some problems of 
bias. The UMLEs of the population sizes were practically unbiased (in terms of the values of their mdre), whereas the HTLEs slightly underestimated 
the population sizes. The performance of the HTLEs of the population totals and means was better in the case of the response variable ``Number of 
friends'' than in the case of the response variable ``Gender''. In the first case, the estimators presented small positive values of r-bias (or mdre), 
whereas in the second case they presented relatively small negative values of r-bias; however, the magnitudes of the values of their r-bias and 
$\sqrt{\textrm{r-mse}}$ were smaller in the first case than in the second one. The opposite situation happened with the HKLEs which presented better
performance in the case of the variable ``Gender'' than in the case of the variable ``Number of friends''. Notice from Figure 1 that the distributions 
of the estimators of the parameters corresponding to $U_1$, that is, $\tau_1$, $Y_1$ and $\bar Y_1$ are approximately symmetrical; those of the parameters 
$\tau_2$ and $Y_2$ are skewed to the right, as well as those of $\hat\tau^{(U)}$ and $\hat{Y}_{HK}^{(U)}$; those of the HTLEs of $\tau$ and 
$Y$ are also approximately symmetrical as well as those of the HKLEs of the means, whereas those of $\hat{\bar Y}_{HT.2}^{(U)}$ and 
$\hat{\bar Y}_{HT}^{(U)}$ are skewed to the left. Notice also that each of these distributions are heavy-tailed as shown by the presence of many outlying 
points on the box plots.

\begin{figure}
	\begin{center}
		\includegraphics[width=5.5in,height=4.0in]{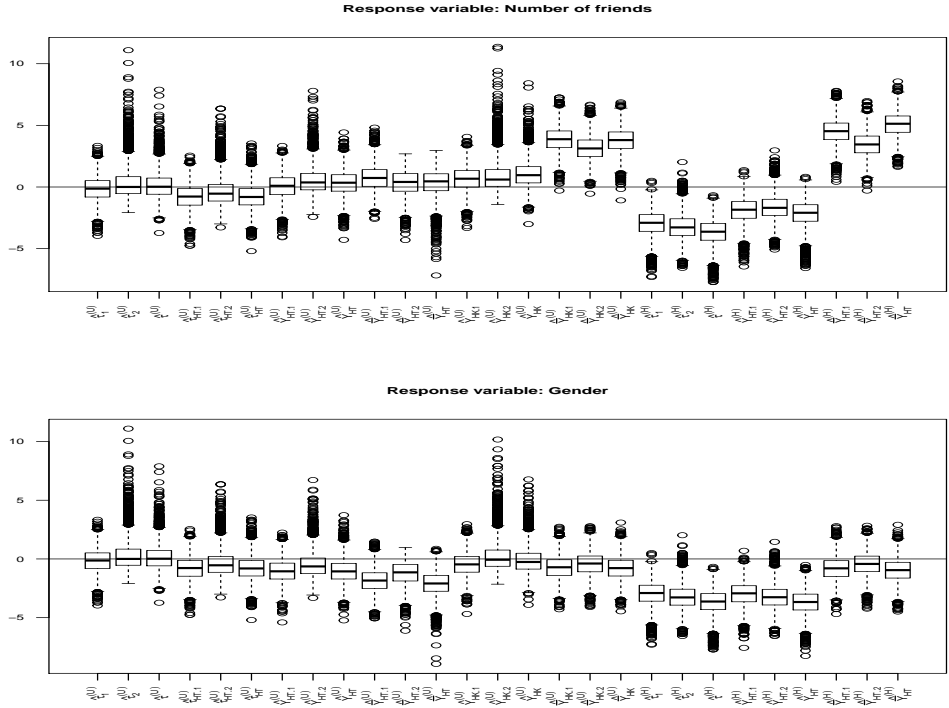}
		\caption{Boxplots for the values of the estimators of the population sizes, totals and means in the population constructed
			using data from the National Longitudinal Study of Adolescent Health.}
	\end{center}
\end{figure}

The performance of the estimators of the population sizes and totals derived under the assumption of homogeneous link probabilities presented problems
of underestimation which produced large values of $\sqrt{\textrm{r-mse}}$ regardless of the type of response variable. However, the estimators of the 
population means performed acceptably, with the exception of the estimator $\hat{\bar Y}_{HT.2}^{(H)}$ which presented a relatively large bias. Notice 
that these estimators of the population means performed pretty well in the case of the binary response variable: very small values of the r-bias and 
$\sqrt{\textrm{r-mse}}$. Finally, notice that the distributions of these estimators are approximately symmetrical and heavy-tailed.

The results of the study about the standard deviation estimations are also shown in Table 1. We can see that every one of the estimators presented problems
of underestimation (in terms of their mdre). The exceptions, perhaps, were the estimators of the standard deviations of the HTLEs 
$\hat{\bar Y}_{HT.1}^{(U)}$, $\hat{\bar Y}_{HT.2}^{(U)}$ and  $\hat{\bar Y}_{HT}^{(U)}$, which in the case of the response variable ``Gender'' presented 
acceptable values of mdre.

{\setlength\tabcolsep{1.5pt}
	\begin{table}
		%\vspace{-.1in}
		\begin{center} 
			\caption{Simulation results obtained for the confidence intervals in a 
				population constructed using data from the National Longitudinal
				Study of Adolescent Health.}
			%			\begin{tabular*}{5.5in}{@{\extracolsep{0.27mm}}
			%					|cc|cccc@{\hspace{1pt}}|cccc@{\hspace{1pt}}|cccc@{\hspace{1pt}}|cccc@{\hspace{1pt}}|}
			\begin{tabular}{|c|c|ccc|ccc|ccc|ccc|}
				\hline
				& & \multicolumn{3}{c|}{CIs based on}    & \multicolumn{3}{c|}{CIs based on}    &
				\multicolumn{3}{c|}{CIs based on}    & \multicolumn{3}{c|}{CIs based on}    \\
				\raisebox{1.0ex}{Response} & & \multicolumn{3}{c|}{UMLEs of sizes}  & \multicolumn{3}{c|}{HTLEs of sizes}  &  
				\multicolumn{3}{c|}{HTLEs of totals}  &  \multicolumn{3}{c|}{HTLEs of means}  \\
				\raisebox{1.0ex}{variable} &  & $\hat{\tau}_{1}^{(U)}$ & $\hat{\tau}_{2}^{(U)}$ & $\hat{\tau}^{(U)}$ &
				$\hat{\tau}_{HT.1}^{(U)}$ & $\hat{\tau}_{HT.2}^{(U)}$ & $\hat{\tau}_{HT}^{(U)}$ &
				$\hat{Y}_{HT.1}^{(U)}$ & $\hat{Y}_{HT.2}^{(U)}$ & $\hat{Y}_{HT}^{(U)}$ &
				$\hat{\bar Y}_{HT.1}^{(U)}$ & $\hat{\bar Y}_{HT.2}^{(U)}$ & $\hat{\bar Y}_{HT}^{(U)}$ \\
				\hline
				Number  & cp   & .89 & .95 & .93 & .82 & .90 & .82 & .90 & .97 & .94 & .81 & .77 & .77 \\
				of      & mrl  & .22 & 1.1 & .36 & .20 & .67 & .24 & .20 & .68 & .23 & .06 & .32 & .15 \\
				friends & mdrl & .21 & .68 & .26 & .19 & .52 & .21 & .20 & .55 & .21 & .06 & .27 & .11 \\
				\hline			
				& cp   & .89 & .95 & .93 & .82 & .90 & .82 & .73 & .85 & .74 & .53 & .93 & .53 \\
				Gender  & mrl  & .22 & 1.1 & .36 & .20 & .67 & .24 & .22 & .66 & .26 & .12 & .33 & .15 \\
				& mdrl & .21 & .68 & .26 & .19 & .52 & .21 & .22 & .54 & .23 & .12 & .29 & .13 \\
				\hline
				\hline
				& & \multicolumn{3}{c|}{CIs based on}     &  \multicolumn{3}{c|}{CIs based on}   &
				\multicolumn{3}{c|}{}  & \multicolumn{3}{c|}{}       \\
				& & \multicolumn{3}{c|}{HKLEs of totals}  &  \multicolumn{3}{c|}{HKLEs of means} &
				\multicolumn{3}{c|}{}  & \multicolumn{3}{c|}{}       \\
				& &  $\hat{Y}_{HK.1}^{(U)}$ & $\hat{Y}_{HK.2}^{(U)}$ & $\hat{Y}_{HK}^{(U)}$ &
				$\hat{\bar Y}_{HK.1}^{(U)}$ & $\hat{\bar Y}_{HK.2}^{(U)}$ & $\hat{\bar Y}_{HK}^{(U)}$ &  
				\multicolumn{3}{c|}{}  & \multicolumn{3}{c|}{}       \\
				\hline				
				Number  & cp   &  .84 & .98 & .88 & .02 & .16 & .09 &  &  &  &  &  &     \\
				of      & mrl  &  .22 & 1.1 & .33 & .05 & .16 & .07 &  &  &  &  &  &     \\
				friends & mdrl &  .22 & .71 & .25 & .05 & .16 & .06 &  &  &  &  &  &     \\
				\hline			
				& cp           &  .84 & .92 & .88 & .85 & .99 & .86 &  &  &  &  &  &     \\
				Gender  & mrl  &  .24 & 1.0 & .37 & .11 & .19 & .10 &  &  &  &  &  &     \\
				& mdrl         &  .23 & .68 & .28 & .11 & .19 & .10 &  &  &  &  &  &     \\				
				\hline
				\hline
				& & \multicolumn{3}{c|}{CIs based on}  &  \multicolumn{3}{c|}{CIs based on}    &
				\multicolumn{3}{c|}{CIs based on}      &  \multicolumn{3}{c|}{}    \\
				& & \multicolumn{3}{c|}{homogeneous}   &  \multicolumn{3}{c|}{homogeneous}     &
				\multicolumn{3}{c|}{homogeneous}       &  \multicolumn{3}{c|}{}     \\				
				& & \multicolumn{3}{c|}{MLEs of sizes} &  \multicolumn{3}{c|}{HTLEs of totals} & 
				\multicolumn{3}{c|}{HTLEs of means}    &  \multicolumn{3}{c|}{}  \\
				&  & $\hat{\tau}_{1}^{(H)}$ & $\hat{\tau}_{2}^{(H)}$ & $\hat{\tau}^{(H)}$ & 
				$\hat{Y}_{HT.1}^{(H)}$ & $\hat{Y}_{HT.2}^{(H)}$ & $\hat{Y}_{HT}^{(H)}$    &
				$\hat{\bar Y}_{HT.1}^{(H)}$ & $\hat{\bar Y}_{HT.2}^{(H)}$ & $\hat{\bar Y}_{HT.1}^{(H)}$ & & & \\
				\hline
				Number  & cp   & .04 & .10 & .01 & .42 & .49 & .28 & .01 & .05 & .00 & & &  \\
				of      & mrl  & .11 & .22 & .10 & .16 & .28 & .14 & .04 & .14 & .05 & & &  \\
				friends & mdrl & .11 & .22 & .10 & .16 & .27 & .14 & .04 & .14 & .05 & & &  \\
				\hline			
				& cp           & .04 & .10 & .01 & .10 & .11 & .02 & .81 & .96 & .77 & & &   \\
				Gender  & mrl  & .11 & .22 & .10 & .17 & .25 & .14 & .10 & .16 & .09 & & &   \\
				& mdrl         & .11 & .22 & .10 & .17 & .25 & .14 & .10 & .16 & .08 & & &   \\
				\hline
				\multicolumn{14}{l}{Notes: Results for confidence intervals derived under the heterogeneity assumption are} \\
				\multicolumn{14}{l}{based on 500 samples and those derived under the homogeneity assumption are based} \\
				\multicolumn{14}{l}{on 5000 samples. Sampling fractions are $f_1=0.46$ and $f_2=0.40$. No convergence} \\
				\multicolumn{14}{l}{problems were observed.}
			\end{tabular}		
		\end{center}
\end{table}}

The results on the 95\% CIs are shown in Table 2. We can see that the CIs of the population sizes based on the UMLEs $\hat\tau^{(U)}_1$,
$\hat\tau^{(U)}_2$ and $\hat\tau^{(U)}$ presented acceptable performance, although the mdrl of the CI of $\tau_2$ was relatively large. The CIs of the 
population sizes based on the HTLEs $\hat\tau^{(U)}_{HT.1}$, $\hat\tau^{(U)}_{HT.2}$ and $\hat\tau^{(U)}_{HT}$ presented relatively low values of the 
cp (between $0.82$ and $0.9$), but acceptable values of their mdrl. The performance of the CIs of the population totals based on the HTLEs 
$\hat Y^{(U)}_{HT.1}$, $\hat Y^{(U)}_{HT.2}$ and $\hat Y^{(U)}_{HT}$ had good performance in the case of the response variable ``Number of friends'': 
both their cp and mdrl presented acceptable values; however, in the case of the response variable ``Gender'' their performance was not good because 
their cp were somehow low: between $0.73$ and $0.85$. The performance of the CIs of the population means was neither good: although their mdrl were 
very short, in the case of the response variable ``Number of friends'' their cp were low (between $0.77$ and $0.81$), and in the case of the variable 
``Gender'' they were very low ($0.53$ in the case of the CIs of $\bar Y_1$ and $\bar Y$, and $0.93$ in the case of the CI of $\bar Y_2$). Nevertheless 
the small cp of the CIs of the population proportions, these CIs still provide valuable information about the proportions. For instance, in the case of 
the CI for $\bar Y_1$, the value of the r-bias of $\hat{\bar Y}_{HT.1}^{(U)}$ was $-0.06$, and since $\bar Y_1=0.466$, it follows that the average value 
of $\hat{\bar Y}_{HT.1}^{(U)}$ was $0.94\times\bar Y_1=0.438$. In addition, since the mrl of the CI was $0.12$, it follows that the average values of the 
lower and upper limits of the CI were $0.410$ and $0.466$, respectively. Consequently, from this point of view, the performance of the CI was acceptable 
despite its small cp. The performance of CIs of the population totals based on the HKLEs was moderately acceptable because of their relatively low cp 
and the relatively large mrl of the CI of $Y_2$. In the case of the CIs of the population means based on this type of estimator, their performance 
was acceptable in the case of the variable ``Gender'', but their cp were very small in the case of the variable ``Number of friends''. However, these CIs 
still provide valuable information because the ``average'' CIs of $\bar Y_1=5.612$, $\bar Y_2=3.915$ and $\bar Y=5.138$ were $(5.75,6.03)$, $(3.93,4.55)$ 
and $(5.35,5.65)$, respectively. Finally, in the case of the CIs constructed under the assumption of homogeneous link probabilities their cp were very low; 
however, their mdrl were very short, and since the values of the r-bias of the estimators of the means were small, we could also conclude that the performance 
of these CIs of the means was still acceptable. Notice also that in the case of the response variable ``Gender'', the cp of the CIs of the means were not so 
low (between $0.77$ and $0.96$).

From the results of this study, we have that inferences based on the UMLEs of the population totals are pretty acceptable, those based on the HTLEs of the
population sizes are acceptable, but not as good as those based on the UMLEs. In the case of the response variable ``Number of friends'',  inferences about
the population totals and means based on the HTLEs are acceptable, although the 95\% CIs of the means tended to have relatively low values of the cp, and 
they are better than those based on the HKLEs. However, in the case of the variable ``Gender'' the opposite happened. This result is consequence of the
higher correlation between the inclusion probabilities and the variable ``Number of friends'' than between the inclusion probabilities and the variable 
``Gender''. (See Thompson, 2002, Ch. 6.)

\subsection{Populations constructed using artificial data}

We constructed two populations whose characteristics are described in Table 3. The difference between the two populations is that in
Population I the link probabilities were generated by using expression $(\ref{rasch})$, that is, under the assumed model, whereas in Population
II they were generated by the following latent-class model used by Pledger (2000) in the context of capture-recapture studies:
$p_{ij}=\exp[\mu^{(k)}+\alpha_i^{(k)}+\beta_j^{(k)}+(\alpha\beta)_{ij}^{(k)}]/\{1+\exp[\mu^{(k)}+\alpha_i^{(k)}+\beta_j^{(k)}+
(\alpha\beta)_{ij}^{(k)}]\}$, $i=1,\ldots,n$; $j=1,2$, and $k=1,2$. In this model, the people in $U_k$ are divided into two classes
according to their propensities to be linked to the sample clusters. The probability that a person in $U_k$ is in class $j$ is
$p_j^{(k)}$ and it is the same for each person in $U_k$. The values of the parameters that appear in each of the two expressions
of the link probabilities were set so that when the size of the initial sample of clusters is $n=15$, in both populations the sampling
fractions were $f_1=0.5$ in $U_1$ and $f_2=0.4$ in $U_2$. Notice from Table 1 that associated with each element of each population there
are two values of two response variables. One variable is a continuous variable whose value associated with the $j$-th element of $U_k$
was obtained by sampling from a non-central chi-square distribution with two degrees of freedom and non-centrality parameter $\psi_j^{(k)}$.
The other variable is a binary variable whose value associated with that element was obtained from a Bernoulli distribution with mean
$\phi_j^{(k)}$. The values of the parameters that appear in the expressions of $\psi_j^{(k)}$ and $\phi_j^{(k)}$ were set so that the values
of the population means of the continuous variable in both populations were $\bar Y_1\approx 50$ and  $\bar Y_2\approx 40$, whereas the
corresponding values of the binary variable were $\bar Y_1\approx 0.3$ and  $\bar Y_2\approx 0.2$. Furthermore, they were set so that for
$n=15$, in Population I the values of the Pearson correlation coefficients between the values of the continuous response variable and
those of the inclusion probabilities were $\rho(y^{(1)},\pi_{(1)})\approx 0.8$ and $\rho(y^{(2)},\pi_{(2)})\approx 0.7$, whereas
the corresponding values for the binary response variable were $\rho(y^{(1)},\pi_{(1)})\approx 0.3$ and $\rho(y_j^{(2)},\pi_{(2)})
\approx 0.27$. In the case of Population II and continuous response variable those values were $\rho(y^{(1)},\pi_{(1)})\approx 0.15$ 
and $\rho(y^{(2)},\pi_{(2)})\approx 0.1$, whereas the corresponding values for the binary variable were $\rho(y^{(1)},\pi_{(1)})
\approx 0$ and $\rho(y^{(2)},\pi_{(2)})\approx 0.1$.

{\setlength\tabcolsep{1.5pt}
	\begin{table}
		%\vspace{-.1in}
		\begin{center} 
			\caption{Parameters of simulated populations.}
			\begin{tabular}{|cccc|cccc|}
				\hline
				\multicolumn{4}{|c|}{Population I} & \multicolumn{4}{c|}{Population II} \\
				\hline
				\multicolumn{4}{|c|}{$N=150$} & \multicolumn{4}{c|}{$N=150$}  \\
				\multicolumn{4}{|c|}{$M_i\sim$ zero trunc. neg. binom. distribution} &
				\multicolumn{4}{c|}{$M_i\sim$ zero trunc. neg. binom. distribution}  \\
				\multicolumn{2}{|c}{$E(M_i)=8$,} & \multicolumn{2}{c|}{$V(M_i)=24$} &
				\multicolumn{2}{c}{$E(M_i)=8$,} & \multicolumn{2}{c|}{$V(M_i)=24$}  \\  
	            \multicolumn{4}{|c|}{$\tau_1=1208,\: \tau_2=400, \: \tau=1608$} & \multicolumn{4}{c|}{$\tau_1=1208,\: \tau_2=400, \: \tau=1608$}  \\							
				\multicolumn{4}{|c|}{$p_{ij}^{(k)}=\frac{\exp\left(\alpha_i^{(k)}+\beta_j^{(k)}\right)}
					{1+\exp\left(\alpha_i^{(k)}+\beta_j^{(k)}\right)}$} &
				\multicolumn{4}{c|}{$p_{ij}^{(k)}=\frac{\exp\left(\mu^{(k)}+\alpha_i^{(k)}+\beta_j^{(k)}+(\alpha\beta)_{ij}^{(k)}\right)}
					{1+\exp\left(\mu^{(k)}+\alpha_i^{(k)}+\beta_j^{(k)}+(\alpha\beta)_{ij}^{(k)}\right)}$}  \\
				\multicolumn{4}{|c|}{$i=1,\ldots,N,\: j=1,\ldots,\tau_k,\: k=1,2$} &  
				\multicolumn{4}{c|}{$i=1,\ldots,N,\: j=1,2,\: k=1,2$}  \\
				\multicolumn{4}{|c|}{} & \multicolumn{2}{c}{$\mu^{(1)}=0.25$,} & \multicolumn{2}{c|}{$\mu^{(2)}=0.05$}  \\
				\multicolumn{4}{|c|}{$\alpha_i^{(k)}=\frac{c_k}{0.001+M_i^{1/4}},\, i=1,\ldots,N,\, k=1,2$} &
				\multicolumn{4}{c|}{$\alpha_i^{(k)}=\frac{c_k}{0.001+M_i^{1/4}},\, i=1,\ldots,N,\, k=1,2$} \\
				\multicolumn{2}{|c}{$c_1=-5.45$,} & \multicolumn{2}{c|}{$c_2=-5.85$} &
				\multicolumn{2}{c}{$c_1=-12$,} & \multicolumn{2}{c|}{$c_2=-12$}  \\
				\multicolumn{4}{|c|}{$\beta_j^{(k)}\sim N(0,1),\: j=1,\ldots,\tau_k$} & \multicolumn{4}{c|}{$\beta_1^{(k)}=1.5,\: \beta_2^{(k)}=0,\: k=1,2$}  \\
				\multicolumn{4}{|c|}{} & \multicolumn{4}{c|}{$(\alpha\beta)_{i1}^{(k)}\sim N(0,1.25^2),\: (\alpha\beta)_{i2}^{(k)}=0,\: k=1,2$}  \\
				\multicolumn{4}{|c|}{} & \multicolumn{4}{c|}{$i=1,\ldots,N$}  \\
				\multicolumn{4}{|c|}{} & \multicolumn{4}{c|}{$p_1^{(k)}=0.3,\: p_2^{(k)}=0.7,\: k=1,2$}  \\
				\multicolumn{4}{|c|}{} &  \multicolumn{4}{c|}{}   \\
				\multicolumn{4}{|c|}{Continuous response variable:} & \multicolumn{4}{c|}{Continuous response variable:}  \\
				\multicolumn{4}{|c|}{$Y_j^{(k)}\sim\chi^2_2(\psi_j^{(k)}),\: j=1,\ldots,\tau_k,\: k=1,2$}  &
				\multicolumn{4}{c|}{$Y_j^{(k)}\sim\chi^2_2(\psi_j^{(k)}),\: j=1,\ldots,\tau_k,\: k=1,2$}  \\
				\multicolumn{4}{|c|}{$\psi^{(k)}_j=5+\frac{d_k\exp(\beta_j^{(k)})}{1+\exp(\beta_j^{(k)})},\:j=1,\ldots,\tau_k,\: k=1,2$} &
				\multicolumn{4}{c|}{$\psi^{(k)}_j=5+\frac{d_k\exp(\mu_k+\beta_j^{(k)})}{1+\exp(\mu_k+\beta_j^{(k)})},\: j=1,\ldots,\tau_k,\: k=1,2$}  \\
				\multicolumn{2}{|c}{$d_1=87$,} & \multicolumn{2}{c|}{$d_2=65$}  & \multicolumn{2}{c}{$d_1=65.05$,} & \multicolumn{2}{c|}{$d_2=50.05$}  \\
				\multicolumn{2}{|c}{$\rho(y^{(1)},\pi_{(1)})=0.79$,}  & \multicolumn{2}{c|}{$\rho(y^{(2)},\pi_{(2)})=0.72$}  &
				\multicolumn{2}{c}{$\rho(y^{(1)},\pi_{(1)})=0.16$,}  & \multicolumn{2}{c|}{$\rho(y^{(2)},\pi_{(2)})=0.11$}  \\
				\multicolumn{4}{|c|}{$Y_1=60390.34,\: Y_2=15945.89, \: Y=76336.22$} & \multicolumn{4}{c|}{$Y_1=60289.03,\: Y_2=16112.78, \: Y=76401.81$}  \\
				\multicolumn{4}{|c|}{$\bar Y_1=49.99,\: \bar Y_2=39.87, \: \bar Y=47.47$} & \multicolumn{4}{c|}{$\bar Y_1=49.91,\: \bar Y_2=40.28, \: \bar Y=47.51$}  \\
				\multicolumn{4}{|c|}{} &  \multicolumn{4}{c|}{}   \\
				\multicolumn{4}{|c|}{Binary response variable:} & \multicolumn{4}{c|}{Binary response variable:}  \\
				\multicolumn{4}{|c|}{$Y_j^{(k)}\sim\textrm{Bernoulli}(\phi_j^{(k)}),\: j=1,\ldots,\tau_k,\: k=1,2$}  &
				\multicolumn{4}{c|}{$Y_j^{(k)}\sim\textrm{Bernoulli}(\phi_j^{(k)}),\: j=1,\ldots,\tau_k,\: k=1,2$}  \\
				\multicolumn{4}{|c|}{$\phi^{(k)}_j=\frac{g_k\exp(\beta_j^{(k)})}{1+\exp(\beta_j^{(k)})},\: j=1,\ldots,\tau_k,\: k=1,2$} &
				\multicolumn{4}{c|}{$\phi^{(k)}_j=\frac{g_k\exp(\mu_k+\beta_j^{(k)})}{1+\exp(\mu_k+\beta_j^{(k)})},\: j=1,\ldots,\tau_k,\: k=1,2$}  \\
				\multicolumn{2}{|c}{$g_1=0.6$,} & \multicolumn{2}{c|}{$g_2=0.39$}  & \multicolumn{2}{c}{$g_1=0.46$,} & \multicolumn{2}{c|}{$g_2=0.33$}  \\
				\multicolumn{2}{|c}{$\rho(y^{(1)},\pi_{(1)})=0.29$,}  & \multicolumn{2}{c|}{$\rho(y^{(2)},\pi_{(2)})=0.27$}  &
				\multicolumn{2}{c}{$\rho(y^{(1)},\pi_{(1)})=-0.01$,}  & \multicolumn{2}{c|}{$\rho(y^{(2)},\pi_{(2)})=0.09$}  \\
				\multicolumn{4}{|c|}{$Y_1=366,\: Y_2=82, \: Y=448$} & \multicolumn{4}{c|}{$Y_1=365,\: Y_2=79, \: Y=444$}  \\
				\multicolumn{4}{|c|}{$\bar Y_1=0.303,\: \bar Y_2=0.205, \: \bar Y=0.279$} & \multicolumn{4}{c|}{$\bar Y_1=0.302,\: \bar Y_2=0.198, \: \bar Y=0.276$}  \\
				\hline
			\end{tabular}		
		\end{center}
\end{table}}

\renewcommand{\baselinestretch}{0.85} 
\small\normalsize

{\setlength\tabcolsep{1.5pt}
	\begin{table}
		%\vspace{-.1in}
		\begin{center} 
			\caption{Relative biases, square roots of relative mean square errors and medians
				of relative errors and absolute relative errors of the estimators of the 
				population sizes, totals and means.}
			%			\begin{tabular*}{5.5in}{@{\extracolsep{0.27mm}}
			%					|cc|cccc@{\hspace{1pt}}|cccc@{\hspace{1pt}}|cccc@{\hspace{1pt}}|cccc@{\hspace{1pt}}|}
			\begin{tabular}{|cc|cccc|cccc|cccc|cccc|}
				\hline
				\multicolumn{2}{|c|}{Population} & \multicolumn{8}{c|}{I} & \multicolumn{8}{c|}{II}  \\ 
				\hline
				\multicolumn{2}{|c|}{Sampling rates} & \multicolumn{4}{c}{$f_{1}=0.5$} & \multicolumn{4}{c|}{$f_{2}=0.4$} &
				\multicolumn{4}{c}{$f_{1}=0.5$} & \multicolumn{4}{c|}{$f_{2}=0.4$} \\ 
				\hline
				\multicolumn{2}{|c|}{Resp. variable} & \multicolumn{4}{c|}{Continuous} & \multicolumn{4}{c|}{Binary} &
				\multicolumn{4}{c|}{Continuous} & \multicolumn{4}{c|}{Binary} \\ 	
				\hline			
				\multicolumn{2}{|c|} {\raisebox{3ex}{Estimator}} & \shortstack{\\r\\b\\i\\a\\s} & 
				$\sqrt{\textrm{\shortstack{r\\m\\s\\e}}}$ & \shortstack{m\\d\\r\\e} & \shortstack{m\\d\\a\\r\\e} &  
				\shortstack{\\r\\b\\i\\a\\s} & $\sqrt{\textrm{\shortstack{r\\m\\s\\e}}}$ & 
				\shortstack{m\\d\\r\\e} & \shortstack{m\\d\\a\\r\\e} &
				\shortstack{\\r\\b\\i\\a\\s} & $\sqrt{\textrm{\shortstack{r\\m\\s\\e}}}$ & 
				\shortstack{m\\d\\r\\e} & \shortstack{m\\d\\a\\r\\e} &
				\shortstack{\\r\\b\\i\\a\\s} & $\sqrt{\textrm{\shortstack{r\\m\\s\\e}}}$ & 
				\shortstack{m\\d\\r\\e} & \shortstack{m\\d\\a\\r\\e} \\
				\hline
				UMLEs & $\hat{\tau}_{1}^{(U)}$ & -$.00$ & $.08$ & -$.01$ & $.05$ & -$.00$ & $.08$ & -$.01$ & $.05$	
				& -$.09$ & $.13$ & -$.08$ & $.09$ & -$.09$ & $.13$ & -$.08$ & $.09$ \\ 
				of & $\hat{\tau}_{2}^{(U)}$ & $.06$ & $.37$ & -$.01$ & $.16$ & $.06$ & $.37$ & -$.01$ & $.16$
				& -$.16$ & $1.9$ & -$.27$ & $.28$ & -$.16$ & $1.9$ & -$.27$ & $.28$ \\ 
				sizes & $\hat{\tau}^{(U)}$ & $.01$ & $.11$ & $.01$ & $.06$ & $.01$ & $.11$ & $.01$ & $.06$
				& -$.10$ & $.48$ & -$.12$ & $.12$ & -$.10$ & $.48$ & -$.12$ & $.12$ \\ \hline
				HTLEs & $\hat{\tau}_{HT.1}^{(U)}$ & -$.11$ & $.13$ & -$.11$ & $.11$ & -$.11$ & $.13$ & -$.11$ & $.11$
				& -$.15$ & $.18$ & -$.15$ & $.15$ & -$.15$ & $.18$ & -$.15$ & $.15$ \\ 
				of & $\hat{\tau}_{HT.2}^{(U)}$ & -$.19$ & $.24$ & -$.21$ & $.21$ & -$.19$ & $.24$ & -$.21$ & $.21$
				& -$.28$ & $.35$ & -$.32$ & $.32$ & -$.28$ & $.35$ & -$.32$ & $.32$ \\ 
				sizes & $\hat{\tau}^{(U)}_{HT}$ & -$.13$ & $.14$ & -$.13$ & $.13$ & -$.13$ & $.14$ & -$.13$ & $.13$ 
				& -$.18$ & $.20$ & -$.18$ & $.18$ & -$.18$ & $.20$ & -$.18$ & $.18$ \\ \hline

				HTLEs & $\hat{Y}_{HT.1}^{(U)}$ & -$.00$ & $.06$ & -$.00$ & $.04$ & $.01$ & $.07$ & $.01$ & $.05$ 
				& -$.14$ & $.17$ & -$.14$ & $.14$ & -$.16$ & $.19$ & -$.16$ & $.16$  \\ 
				of & $\hat{Y}_{HT.2}^{(U)}$ & -$.06$ & $.17$ & -$.08$ & $.12$ & $.03$ & $.19$ & $.01$ & $.11$ 
				& -$.27$ & $.34$ & -$.30$ & $.31$ & -$.21$ & $.32$ & -$.25$ & $.26$ \\ 
				totals & $\hat{Y}_{HT}^{(U)}$ & -$.01$ & $.06$ & -$.01$ & $.04$ & $.02$ & $.07$ & $.02$ & $.05$ 
				& -$.17$ & $.18$ & -$.17$ & $.17$ & -$.16$ & $.19$ & -$.17$ & $.17$ \\ \hline
				HTLEs & $\hat{\bar Y}_{HT.1}^{(U)}$ & $.00$ & $.03$ & $.00$ & $.02$ & $.02$ & $.05$ & $.02$ & $.03$
				& -$.06$ & $.07$ & -$.07$ & $.07$ & -$.08$ & $.09$ & -$.08$ & $.08$  \\ 
				of & $\hat{\bar Y}_{HT.2}^{(U)}$ & -$.08$ & $.14$ & -$.07$ & $.08$ & $.01$ & $.16$ & $.01$ & $.10$
				& -$.06$ & .$10$ & -$.05$ & $.05$ & $.02$ & $.14$ & $.03$ & $.09$ \\ 
				means & $\hat{\bar Y}_{HT}^{(U)}$ & -$.02$ & $.08$ & -$.01$ & $.03$ & $.01$ & $.09$ & $.01$ & $.04$
				& -$.06$ & $.11$ & -$.06$ & $.06$ & -$.06$ & $.15$ & -$.06$ & $.06$ \\		
				\hline
				
				HKLEs & $\hat{Y}_{HK.1}^{(U)}$ & $.11$ & $.14$ & $.11$ & $.11$ & $.13$ & $.16$ & $.13$ & $.13$ 
				& -$.07$ & $.12$ & -$.07$ & $.08$ & -$.09$ & $.14$ & -$.09$ & $.10$  \\ 
				of & $\hat{Y}_{HK.2}^{(U)}$ & $.24$ & $.47$ & $.15$ & $.18$ & $.35$ & $.57$ & $.25$ & $.25$ 
				& -$.15$ & $1.9$ & -$.26$ & $.27$ & -$.06$ & $2.7$ & -$.20$ & $.22$ \\ 
				totals & $\hat{Y}_{HK}^{(U)}$ & $.15$ & $.18$ & $.14$ & $.14$ & $.18$ & $.21$ & $.17$ & $.17$ 
				& -$.08$ & $.45$ & -$.09$ & $.10$ & -$.08$ & $.49$ & -$.09$ & $.10$ \\ \hline
				HKLEs & $\hat{\bar Y}_{HK.1}^{(U)}$ & $.12$ & $.12$ & $.12$ & $.12$ & $.14$ & $.14$ & $.14$ & $.14$
				& $.02$ & $.02$ & $.02$ & $.02$ & -$.01$ & $.04$ & -$.01$ & $.03$  \\ 
				of & $\hat{\bar Y}_{HK.2}^{(U)}$ & $.17$ & $.17$ & $.17$ & $.17$ & $.27$ & $.30$ & $.27$ & $.27$
				& $.02$ & .$03$ & $.02$ & $.02$ & $.11$ & $.16$ & $.11$ & $.11$ \\ 
				means & $\hat{\bar Y}_{HK}^{(U)}$ & $.13$ & $.13$ & $.13$ & $.13$ & $.17$ & $.17$ & $.17$ & $.17$
				& $.02$ & $.03$ & $.02$ & $.02$ & $.03$ & $.05$ & $.03$ & $.03$ \\	
				
				\hline			
				Homo. & $\hat{\tau}_{1}^{(H)}$ & -$.31$ & $.31$ & -$.31$ & $.31$ & -$.31$ & $.31$ & -$.31$ & $.31$	
				& -$.25$ & $.27$ & -$.26$ & $.26$ & -$.25$ & $.27$ & -$.26$ & $.26$ \\ 
				MLEs & $\hat{\tau}_{2}^{(H)}$ & -$.40$ & $.41$ & -$.40$ & $.40$ & -$.40$ & $.41$ & -$.40$ & $.40$
				& -$.35$ & $.36$ & -$.37$ & $.37$ & -$.35$ & $.36$ & -$.37$ & $.37$ \\ 
				sizes & $\hat{\tau}^{(H)}$ & -$.33$ & $.33$ & -$.33$ & $.33$ & -$.33$ & $.33$ & -$.33$ & $.33$ 
				& -$.27$ & $.29$ & -$.28$ & $.28$  & -$.27$ & $.29$ & -$.28$ & $.28$ \\ \hline 
				Homo. & $\hat{Y}_{HT.1}^{(H)}$ & -$.19$ & $.19$ & -$.19$ & $.19$ & -$.17$ & $.17$ & -$.17$ & $.17$ 
				& -$.24$ & $.25$ & -$.24$ & $.24$ & -$.26$ & $.28$ & -$.26$ & $.26$  \\ 
				HTLEs & $\hat{Y}_{HT.2}^{(H)}$ & -$.29$ & $.29$ & -$.29$ & $.29$ & -$.21$ & $.23$ & -$.21$ & $.21$ 
				& -$.33$ & $.35$ & -$.35$ & $.35$ & -$.27$ & $.31$ & -$.29$ & $.30$ \\ 
				totals & $\hat{Y}_{HT}^{(H)}$ & -$.21$ & $.21$ & -$.21$ & $.21$ & -$.18$ & $.18$ & -$.18$ & $.18$ 
				& -$.26$ & $.27$ & -$.26$ & $.26$ & -$.26$ & $.27$ & -$.26$ & $.26$ \\ \hline 
				Homo. & $\hat{\bar Y}_{HT.1}^{(H)}$ & $.17$ & $.17$ & $.17$ & $.17$ & $.20$ & $.21$ & $.20$ & $.20$ 
				& $.02$ & $.02$ & $.02$ & $.02$ & -$.01$ & $.04$ & -$.01$ & $.03$  \\ 
				HTLEs & $\hat{\bar Y}_{HT.2}^{(H)}$ & $.20$ & $.20$ & $.20$ & $.20$ & $.33$ & $.35$ & $.32$ & $.32$
				& $.02$ & .$03$ & $.02$ & $.02$ & $.12$ & $.16$ & $.12$ & $.12$ \\ 
				means & $\hat{\bar Y}_{HT}^{(H)}$ & $.18$ & $.12$ & $.18$ & $.18$ & $.24$ & $.14$ & $.23$ & $.23$
				& $.03$ & $.03$ & $.03$ & $.03$ & $.02$ & $.08$ & $.02$ & $.03$ \\
				\hline				
				%\multicolumn{10}{l}{Notes: $\hat\tau_k$, and $\hat\tau$, MLEs derived under the 
				% heterogeneity assumption; $\tilde\tau_k$ and} \\
				\multicolumn{18}{l}{Notes: Results are based on 5000 samples. In Population I the percentages of samples in} \\
				\multicolumn{18}{l}{which the estimators derived under the heterogeneity assumption of the population param-} \\
				\multicolumn{18}{l}{eters associated with $U_1$, $U_2$ and $U$ were not computed because of numerical convergence} \\
				\multicolumn{18}{l}{problems were 0\%, 0.02\% and 0.02\%, respectively, whereas in Population II the corres-} \\
				\multicolumn{18}{l}{ponding percentages were 0.36\%, 8.7\% and 9.0\%. In the computation of the estimators} \\ 
				\multicolumn{18}{l}{derived under the homogeneity assumption no convergence problems were presented.} \\
				%			\end{tabular*}
			\end{tabular}		
		\end{center}
\end{table}}

\renewcommand{\baselinestretch}{1.0} 
\small\normalsize

\begin{figure}
	\begin{center}
		\includegraphics[width=5.5in,height=7.0in]{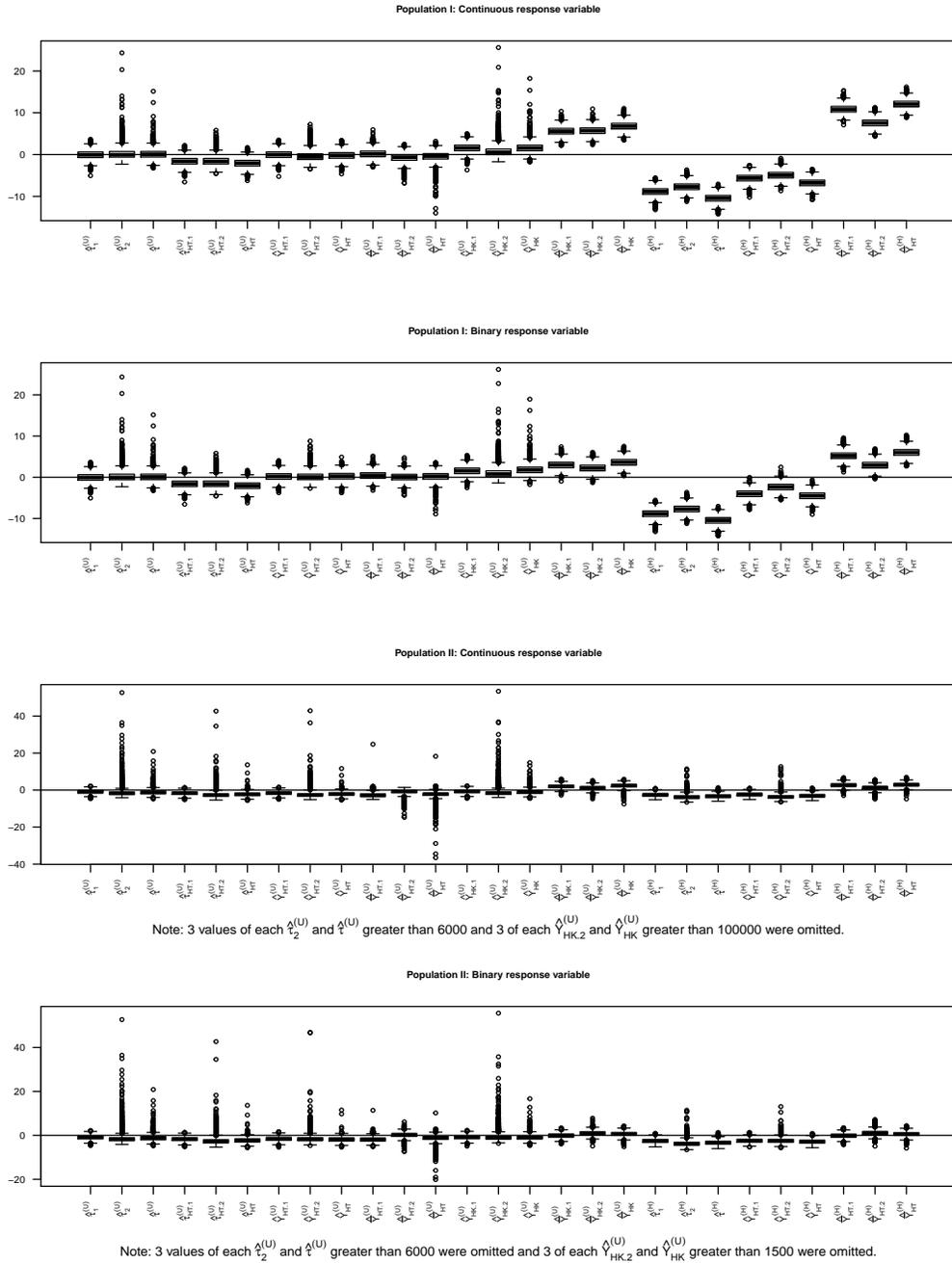}
		\caption{Boxplots for the values of the estimators of the population sizes, totals and means in Populations I and II.}
	\end{center}
\end{figure}

The Monte Carlo study was carried out as in the previous study, except that the size of the initial sample of clusters was $n=15$ and that
for each selected value $m_i$, the value $x_{ij}^{(k)}$ of the link indicator variable $X_{ij}^{(k)}$ was generated from the Bernoulli 
distribution with mean $p_{ij}^{(k)}$ (see its expression in Table 3). 

The results of the study on the point estimators are shown in Table 4 and in Figure 2. We can see that in the case of Population I, the 
performance of the UMLEs of the population sizes was acceptable: they did not show problems of bias and only the estimator $\hat\tau_2^{(U)}$ 
presented problems of instability. The distribution of the estimator $\hat\tau_1^{(U)}$ was symmetrical, whereas those of the estimators 
$\hat\tau_2^{(U)}$ and $\hat\tau^{(U)}$ were skewed to the right with long tails. For that reason the values of their $\sqrt{\textrm{r-mse}}$ 
were greater than those of their mdare. On the other hand, the performance of the HTLEs $\hat\tau_{HT.1}^{(U)}$, $\hat\tau_{HT.2}^{(U)}$ and 
$\hat\tau_{HT}^{(U)}$ was not good because they presented relatively large negative biases. The performance of the HTLEs $\hat Y_{HT.1}^{(U)}$, 
$\hat Y_{HT.2}^{(U)}$ and $\hat Y_{HT}^{(U)}$ of the population totals and that of the HTLEs $\hat{\bar Y}_{HT.1}^{(U)}$, 
$\hat{\bar Y}_{HT.2}^{(U)}$ and $\hat{\bar Y}_{HT}^{(U)}$ of the population means was acceptable in both cases continuous and binary response 
variables: they did not present problems of bias nor problems of instability. It is worth noting that in the case of the continuous response 
variable the distributions of $\hat Y_{HT.1}^{(U)}$ and $\hat Y_{HT}^{(U)}$ were relatively symmetrical; those of $\hat Y_{HT.2}^{(U)}$ and 
$\hat{\bar Y}_{HT.1}^{(U)}$ were somewhat skewed to the right, and those of $\hat{\bar Y}_{HT.2}^{(U)}$ and $\hat{\bar Y}_{HT}^{(U)}$ were 
skewed to the left, whereas in the case of the binary response variable the distribution of every one of the estimators was symmetrical, except 
those of $\hat Y_{HT.2}^{(U)}$ and $\hat{\bar Y}_{HT}^{(U)}$ which were skewed to the right and to the left, respectively. The performance of
the HKLEs of the population totals and means was not good because they presented relatively large positive biases. Finally, the performance of 
the estimators derived under the assumption of homogeneous link probabilities was not good: the estimators of the population sizes and population 
totals presented serious problems of underestimation, whereas those of the population means serious problems of overestimation.

\renewcommand{\baselinestretch}{0.7} 
\small\normalsize

{\setlength\tabcolsep{0.7pt}
	\begin{table}
		\vspace{-.1in}
		\begin{center} 
			\caption{Relative biases, square roots of relative mean square errors and medians
				of relative errors and absolute relative errors of the deviation standard  
				estimators of the estimators of the population sizes, totals and means.}
			%			\begin{tabular*}{5.5in}{@{\extracolsep{0.27mm}}
			%					|cc|cccc@{\hspace{1pt}}|cccc@{\hspace{1pt}}|cccc@{\hspace{1pt}}|cccc@{\hspace{1pt}}|}
			\begin{tabular}{|cc|cccc|cccc|cccc|cccc|}
				\hline
				\multicolumn{2}{|c|}{Population} & \multicolumn{8}{c|}{I} & \multicolumn{8}{c|}{II}  \\ 
				\hline
				\multicolumn{2}{|c|}{Sampling rates} & \multicolumn{4}{c}{$f_{1}=0.5$} & \multicolumn{4}{c|}{$f_{2}=0.4$} &
				\multicolumn{4}{c}{$f_{1}=0.5$} & \multicolumn{4}{c|}{$f_{2}=0.4$} \\ 
				\hline
				\multicolumn{2}{|c|}{Response variable} & \multicolumn{4}{c|}{Continuous}  & \multicolumn{4}{c|}{Binary} &
				\multicolumn{4}{c|}{Continuous} &  \multicolumn{4}{c|}{Binary} \\ 	
				\hline			
				\multicolumn{2}{|c|}{\raisebox{1ex}{ \shortstack{\\Deviation\\standard\\estimator} } }
				& \shortstack{\\r\\b\\i\\a\\s} & 
				$\sqrt{\textrm{\shortstack{r\\m\\s\\e}}}$ & \shortstack{m\\d\\r\\e} & \shortstack{m\\d\\a\\r\\e} &  
				\shortstack{\\r\\b\\i\\a\\s} & $\sqrt{\textrm{\shortstack{r\\m\\s\\e}}}$ & 
				\shortstack{m\\d\\r\\e} & \shortstack{m\\d\\a\\r\\e} &
				\shortstack{\\r\\b\\i\\a\\s} & $\sqrt{\textrm{\shortstack{r\\m\\s\\e}}}$ & 
				\shortstack{m\\d\\r\\e} & \shortstack{m\\d\\a\\r\\e} &
				\shortstack{\\r\\b\\i\\a\\s} & $\sqrt{\textrm{\shortstack{r\\m\\s\\e}}}$ & 
				\shortstack{m\\d\\r\\e} & \shortstack{m\\d\\a\\r\\e} \\
				\hline
				UMLEs & $\widehat{sd}_B(\hat{\tau}_{1}^{(U)})$ & $.15$ & $.29$ & $.12$ & $.16$ & $.15$ & $.29$ & $.12$ & $.16$	
				& -$.19$ & $.27$ & -$.20$ & $.21$ & -$.19$ & $.27$ & -$.20$ & $.21$ \\ 
				of & $\widehat{sd}_B(\hat{\tau}_{2}^{(U)})$ & $2.0$ & $6.7$ & $.22$ & $.58$ & $2.0$ & $6.7$ & $.22$ & $.58$
				& $.26$ & $10.5$ & -$.59$ & $.64$ & $.26$ & $10.5$ & -$.59$ & $.64$ \\ 
				sizes & $\widehat{sd}_B(\hat{\tau}^{(U)})$ & $1.8$ & $5.6$ & $.35$ & $.39$  & $1.8$ & $5.6$ & $.35$ & $.39$
				& $.23$ & $7.3$ & -$.30$ & $.33$ & $.23$ & $7.3$ & -$.30$ & $.33$ \\ \hline
				HTLEs & $\widehat{sd}_B(\hat{\tau}_{HT.1}^{(U)})$ & $.15$ & $.29$ & $.11$ & $.15$ & $.15$ & $.29$ & $.11$ & $.15$
				& -$.30$ & $.35$ & -$.31$ & $.31$ & -$.30$ & $.35$ & -$.31$ & $.31$ \\ 
				of & $\widehat{sd}_B(\hat{\tau}_{HT.2}^{(U)})$ & $.85$ & $1.6$ & $.42$ & $.48$ & $.85$ & $1.6$ & $.42$ & $.48$
				& -$.11$ & $1.9$ & -$.48$ & $.55$ & -$.11$ & $1.9$ & -$.48$ & $.55$ \\ 
				sizes & $\widehat{sd}_B(\hat{\tau}^{(U)}_{HT})$ & $.55$ & $.90$ & $.36$ & $.36$ & $.55$ & $.90$ & $.36$ & $.36$ 
				& -$.18$ & $.94$ & -$.32$ & $.35$ & -$.18$ & $.94$ & -$.32$ & $.35$ \\ \hline
				HTLEs & $\widehat{sd}_B(\hat{Y}_{HT.1}^{(U)})$ & $.20$ & $.32$ & $.16$ & $.19$ & $.12$ & $.26$ & $.10$ & $.15$ 
				& -$.29$ & $.35$ & -$.30$ & $.30$ & -$.23$ & $.30$ & -$.24$ & $.25$  \\ 
				of & $\widehat{sd}_B(\hat{Y}_{HT.2}^{(U)})$ & $.78$ & $1.5$ & $.33$ & $.45$ & $.63$ & $1.3$ & $.26$ & $.41$ 
				& -$.09$ & $1.9$ & -$.47$ & $.54$ & $.02$ & $1.6$ & -$.35$ & $.44$ \\ 
				totals & $\widehat{sd}_B(\hat{Y}_{HT}^{(U)})$ & $.57$ & $.89$ & $.40$ & $.40$ & $.46$ & $.72$ & $.34$ & $.34$ 
				& -$.18$ & $.82$ & -$.31$ & $.34$ & -$.13$ & $.59$ & -$.22$ & $.26$ \\ \hline
				HTLEs & $\widehat{sd}_B(\hat{\bar Y}_{HT.1}^{(U)})$ & $.01$ & $.18$ & -$.01$ & $.11$ & $.16$ & $.23$ & $.15$ & $.16$
				& -$.50$ & $.54$ & -$.51$ & $.51$ & -$.11$ & $.27$ & -$.13$ & $.17$  \\ 
				of & $\widehat{sd}_B(\hat{\bar Y}_{HT.2}^{(U)})$ & $.35$ & $.65$ & $.22$ & $.35$ & $.24$ & $.44$ & $.17$ & $.21$
				& -$.20$ & .$67$ & -$.38$ & $.44$ & $.06$ & $.50$ & -$.06$ & $.24$ \\ 
				means & $\widehat{sd}_B(\hat{\bar Y}_{HT}^{(U)})$ & $.84$ & $1.8$ & $.20$ & $.43$ & $.65$ & $1.3$ & $.19$ & $.28$
				& -$.35$ & $.73$ & -$.50$ & $.52$ & -$.02$ & $.60$ & -$.16$ & $.22$ \\	\hline	
				
				HKLEs & $\widehat{sd}_B(\hat{Y}_{HK.1}^{(U)})$ & $.20$ & $.32$ & $.16$ & $.19$ & $.12$ & $.26$ & $.09$ & $.15$ 
				& -$.18$ & $.27$ & -$.19$ & $.21$ & -$.16$ & $.26$ & -$.16$ & $.19$  \\ 
				of & $\widehat{sd}_B(\hat{Y}_{HK.2}^{(U)})$ & $1.9$ & $6.3$ & $.14$ & $.56$ & $1.6$ & $5.9$ & $.09$ & $.52$ 
				& $.26$ & $10.3$ & -$.59$ & $.63$ & $.32$ & $9.6$ & -$.50$ & $.57$ \\ 
				totals & $\widehat{sd}_B(\hat{Y}_{HK}^{(U)})$ & $1.9$ & $5.6$ & $.34$ & $.37$ & $1.7$ & $5.2$ & $.33$ & $.37$ 
				& $.22$ & $6.7$ & -$.27$ & $.30$ & $.18$ & $5.2$ & -$.21$ & $.26$ \\ \hline
				HKLEs & $\widehat{sd}_B(\hat{\bar Y}_{HK.1}^{(U)})$ & -$.12$ & $.18$ & -$.13$ & $.14$ & $.11$ & $.18$ & $.10$ & $.12$
				& -$.10$ & $.20$ & -$.12$ & $.15$ & $.05$ & $.23$ & $.04$ & $.15$  \\ 
				of & $\widehat{sd}_B(\hat{\bar Y}_{HK.2}^{(U)})$ & $.09$ & $.23$ & $.06$ & $.15$ & $.11$ & $.25$ & $.09$ & $.15$
				& $.07$ & .$35$ & $.00$ & $.20$ & $.03$ & $.33$ & -$.03$ & $.20$ \\ 
				means & $\widehat{sd}_B(\hat{\bar Y}_{HK}^{(U)})$ & $.06$ & $.30$ & -$.01$ & $.14$ & $.16$ & $.29$ & $.13$ & $.14$
				& -$.13$ & $.47$ & -$.24$ & $.29$ & $.08$ & $.32$ & $.06$ & $.16$ \\	\hline

				Homo. & $\widehat{sd}(\hat{\tau}_{1}^{(H)})$ & -$.42$ & $.42$ & -$.42$ & $.42$ & -$.42$ & $.42$ & -$.42$ & $.42$	
				& -$.74$ & $.75$ & -$.75$ & $.75$ & -$.74$ & $.75$ & -$.75$ & $.75$ \\ 
				MLEs & $\widehat{sd}(\hat{\tau}_{2}^{(H)})$ & -$.20$ & $.26$ & -$.22$ & $.23$ & -$.20$ & $.26$ & -$.22$ & $.23$
				& -$.48$ & $.66$ & -$.59$ & $.60$ & -$.48$ & $.66$ & -$.59$ & $.60$ \\ 
				sizes & $\widehat{sd}(\hat{\tau}^{(H)})$ & -$.41$ & $.42$ & -$.42$ & $.42$ & -$.41$ & $.42$ & -$.42$ & $.42$
				& -$.69$ & $.71$ & -$.71$ & $.71$ & -$.69$ & $.71$ & -$.71$ & $.71$ \\ \hline
				Homo. & $\widehat{sd}(\hat{Y}_{HT.1}^{(H)})$ & -$.24$ & $.25$ & -$.24$ & $.24$ & -$.18$ & $.19$ & -$.18$ & $.18$ 
				& -$.69$ & $.71$ & -$.72$ & $.72$ & -$.62$ & $.64$ & -$.65$ & $.65$  \\ 
				HTLEs & $\widehat{sd}(\hat{Y}_{HT.2}^{(H)})$ & -$.11$ & $.21$ & -$.14$ & $.17$ & -$.08$ & $.19$ & -$.10$ & $.14$ 
				& -$.47$ & $.65$ & -$.58$ & $.59$ & -$.31$ & $.54$ & -$.42$ & $.45$ \\ 
				totals & $\widehat{sd}(\hat{Y}_{HT}^{(H)})$ & -$.25$ & $.27$ & -$.26$ & $.26$ & -$.17$ & $.19$ & -$.18$ & $.18$ 
				& -$.66$ & $.68$ & -$.69$ & $.69$ & -$.58$ & $.60$ & -$.61$ & $.61$ \\ \hline
				Homo. & $\widehat{sd}(\hat{\bar Y}_{HT.1}^{(H)})$ & -$.38$ & $.39$ & -$.39$ & $.39$ & -$.16$ & $.18$ & -$.16$ & $.16$
				& -$.21$ & $.29$ & -$.23$ & $.24$ & -$.09$ & $.25$ & -$.11$ & $.18$  \\ 
				HTLEs & $\widehat{sd}(\hat{\bar Y}_{HT.2}^{(H)})$ & -$.19$ & $.22$ & -$.20$ & $.20$ & -$.11$ & $.15$ & -$.12$ & $.13$
				& -$.12$ & .$29$ & -$.16$ & $.22$ & -$.12$ & $.29$ & -$.15$ & $.22$ \\ 
				means & $\widehat{sd}(\hat{\bar Y}_{HT}^{(H)})$ & -$.37$ & $.38$ & -$.38$ & $.38$ & -$.16$ & $.18$ & -$.17$ & $.17$
				& -$.26$ & $.36$ & -$.30$ & $.31$ & -$.13$ & $.25$ & -$.15$ & $.19$ \\
				\hline				
				\multicolumn{18}{l}{Notes: Bootstrap standard deviation estimators $\widehat{sd}_B$ were computed using 50 bootstrap} \\ 
				\multicolumn{18}{l}{samples and their results are based on 500 replicated samples. In Population II the per-} \\					
				\multicolumn{18}{l}{centages of replicated samples in which the estimators $\widehat{sd}_B$ were not computed because} \\
				\multicolumn{18}{l}{of convergence problems were 0.8\%, 8.2\% and 8.8\% in $U_1$, $U_2$ and $U$, respectively,} \\
				\multicolumn{18}{l}{whereas in Population I the respective percentages were all 0\%. Results on sd estima-} \\
				\multicolumn{18}{l}{tors derived under the homogeneity assumption are based on 5000 replicated samples} \\
				\multicolumn{18}{l}{ and no convergence problems were presented.}
				%			\end{tabular*}
			\end{tabular}		
		\end{center}
\end{table}}

\renewcommand{\baselinestretch}{0.98} 
\small\normalsize

{\setlength\tabcolsep{5.0pt}
	\begin{table}
		\vspace{-.1in}
		\begin{center} 
			\caption{Coverage probabilities and means and medians of relative lengths of
				the 95\% confidence intervals of the population sizes, totals and means.}
			%			\begin{tabular*}{5.5in}{@{\extracolsep{0.27mm}}
			%					|cc|cccc@{\hspace{1pt}}|cccc@{\hspace{1pt}}|cccc@{\hspace{1pt}}|cccc@{\hspace{1pt}}|}
			\begin{tabular}{|cc|ccc|ccc|ccc|ccc|}
				\hline
				\multicolumn{2}{|c|}{Population} & \multicolumn{6}{c|}{I} & \multicolumn{6}{c|}{II}  \\ 
				\hline
				\multicolumn{2}{|c|}{Sampling rates} & \multicolumn{3}{c}{$f_{1}=0.5$} & \multicolumn{3}{c|}{$f_{2}=0.4$} &
				\multicolumn{3}{c}{$f_{1}=0.5$} & \multicolumn{3}{c|}{$f_{2}=0.4$} \\ 
				\hline
				\multicolumn{2}{|c|}{Response variable} & \multicolumn{3}{c|}{Continuous} & \multicolumn{3}{c|}{Binary} &
				\multicolumn{3}{c|}{Continuous} & \multicolumn{3}{c|}{Binary} \\ 	
				\hline			
				\multicolumn{2}{|c|}{95\% CI} & cp & mrl & mdrl & cp & mrl & mdrl & cp & mrl & mdrl & cp & mrl & mdrl \\
				\hline
				UMLEs & $\textrm{CI}(\hat{\tau}_{1}^{(U)})$ & $.95$ & $.37$ & $.36$ & $.95$ & $.37$ & $.36$ 
				& $.76$ & $.29$ & $.29$ & $.76$ & $.29$ & $.29$ \\ 
				of & $\textrm{CI}(\hat{\tau}_{2}^{(U)})$ & $.97$ & $5.6$ & $1.8$ & $.97$ & $5.6$ & $1.8$ 
				& $.51$ & $1.7$ & $.44$ & $.51$ & $1.7$ & $.44$ \\ 
				sizes & $\textrm{CI}(\hat{\tau}^{(U)})$ & $.98$ & $1.4$ & $.52$ & $.98$ & $1.4$ & $.52$ 
				& $.63$ & $.54$ & $.26$ & $.63$ & $.54$ & $.26$ \\ \hline
				HTLEs & $\textrm{CI}(\hat{\tau}_{HT.1}^{(U)})$ & $.78$ & $.31$ & $.30$ & $.78$ & $.31$ & $.30$
				& $.44$ & $.24$ & $.24$ & $.44$ & $.24$ & $.24$ \\ 
				of & $\textrm{CI}(\hat{\tau}_{HT.2}^{(U)})$ & $.85$ & $1.2$ & $.85$ & $.85$ & $1.2$ & $.85$
				& $.36$ & $.66$ & $.34$ & $.36$ & $.66$ & $.34$ \\ 
				sizes & $\textrm{CI}(\hat{\tau}^{(U)}_{HT})$ & $.78$ & $.39$ & $.33$ & $.78$ & $.39$ & $.33$
				& $.28$ & $.27$ & $.21$ & $.28$ & $.27$ & $.21$ \\ \hline
				
				HTLEs & $\textrm{CI}(\hat{Y}_{HT.1}^{(U)})$ & $.96$ & $.29$ & $.28$ & $.96$ & $.32$ & $.32$ 
				& $.42$ & $.24$ & $.23$ & $.53$ & $.29$ & $.28$ \\ 
				of & $\textrm{CI}(\hat{Y}_{HT.2}^{(U)})$ & $.90$ & $1.1$ & $.81$ & $.98$ & $1.4$ & $1.0$ 
				& $.28$ & $.56$ & $.33$ & $.63$ & $.82$ & $.49$ \\ 
				totals & $\textrm{CI}(\hat{Y}_{HT}^{(U)})$ & $.96$ & $.35$ & $.31$ & $.98$ & $.38$ & $.35$ 
				& $.27$ & $.25$ & $.21$ & $.47$ & $.30$ & $.26$ \\ \hline
				HTLEs & $\textrm{CI}(\hat{\bar Y}_{HT.1}^{(U)})$ & $.95$ & $.10$ & $.09$ & $.96$ & $.19$ & $.19$ 
				& $.19$ & $.08$ & $.07$ & $.55$ & $.18$ & $.18$ \\ 
				of & $\textrm{CI}(\hat{\bar Y}_{HT.2}^{(U)})$ & $.98$ & $.57$ & $.52$ & $.96$ & $.79$ & $.75$ 
				& $.91$ & $.23$ & $.18$ & $.95$ & $.58$ & $.51$ \\ 
				means & $\textrm{CI}(\hat{\bar Y}_{HT}^{(U)})$ & $.98$ & $.33$ & $.21$ & $.96$ & $.41$ & $.29$ 
				& $.38$ & $.11$ & $.08$ & $.84$ & $.22$ & $.19$ \\	\hline	
								
				HKLEs & $\textrm{CI}(\hat{Y}_{HK.1}^{(U)})$ & $.80$ & $.34$ & $.33$ & $.66$ & $.38$ & $.37$ 
				& $.74$ & $.28$ & $.28$ & $.78$ & $.33$ & $.33$ \\ 
				of & $\textrm{CI}(\hat{Y}_{HK.2}^{(U)})$ & $.98$ & $4.1$ & $1.6$ & $.99$ & $6.4$ & $2.0$ 
				& $.45$ & $1.3$ & $.43$ & $.75$ & $1.9$ & $.60$ \\ 
				totals & $\textrm{CI}(\hat{Y}_{HK}^{(U)})$ & $.94$ & $1.1$ & $.50$ & $.79$ & $1.3$ & $.54$ 
				& $.64$ & $.44$ & $.26$ & $.75$ & $.53$ & $.31$ \\ \hline
				HKLEs & $\textrm{CI}(\hat{\bar Y}_{HK.1}^{(U)})$ & $.00$ & $.07$ & $.07$ & $.17$ & $.19$ & $.19$ 
				& $.48$ & $.03$ & $.03$ & $.96$ & $.17$ & $.17$ \\ 
				of & $\textrm{CI}(\hat{\bar Y}_{HK.2}^{(U)})$ & $.00$ & $.13$ & $.12$ & $.48$ & $.55$ & $.54$ 
				& $.77$ & $.07$ & $.07$ & $.84$ & $.47$ & $.44$ \\ 
				means & $\textrm{CI}(\hat{\bar Y}_{HK}^{(U)})$ & $.00$ & $.08$ & $.08$ & $.08$ & $.21$ & $.20$ 
				& $.30$ & $.04$ & $.04$ & $.92$ & $.18$ & $.17$ \\	\hline	
										
				Homo. & $\textrm{CI}(\hat{\tau}_{1}^{(H)})$ & $.00$ & $.08$ & $.08$ & $.00$ & $.08$ & $.08$ 
				& $.05$ & $.10$ & $.10$ & $.05$ & $.10$ & $.10$ \\ 
				MLEs & $\textrm{CI}(\hat{\tau}_{2}^{(H)})$ & $.00$ & $.17$ & $.16$ & $.00$ & $.17$ & $.16$ 
				& $.09$ & $.24$ & $.19$ & $.09$ & $.24$ & $.19$ \\ 
				sizes & $\textrm{CI}(\hat{\tau}^{(H)})$ & $.00$ & $.07$ & $.07$ & $.00$ & $.07$ & $.07$ 
				& $.02$ & $.10$ & $.09$ & $.02$ & $.10$ & $.09$ \\ \hline
				Homo. & $\textrm{CI}(\hat{Y}_{HT.1}^{(H)})$ & $.00$ & $.10$ & $.10$ & $.01$ & $.14$ & $.14$ 
				& $.09$ & $.12$ & $.11$ & $.10$ & $.15$ & $.14$  \\ 
				HTLEs & $\textrm{CI}(\hat{Y}_{HT.2}^{(H)})$ & $.00$ & $.21$ & $.20$ & $.32$ & $.32$ & $.32$ 
				& $.10$ & $.25$ & $.20$ & $.26$ & $.38$ & $.32$ \\ 
				totals & $\textrm{CI}(\hat{Y}_{HT}^{(H)})$ & $.00$ & $.09$ & $.09$ & $.00$ & $.13$ & $.13$ 
				& $.03$ & $.11$ & $.10$ & $.06$ & $.15$ & $.14$ \\ \hline
				Homo. & $\textrm{CI}(\hat{\bar Y}_{HT.1}^{(H)})$ & $.00$ & $.04$ & $.04$ & $.00$ & $.13$ & $.13$ 
				& $.17$ & $.03$ & $.03$ & $.91$ & $.14$ & $.13$  \\ 
				HTLEs & $\textrm{CI}(\hat{\bar Y}_{HT.2}^{(H)})$ & $.00$ & $.08$ & $.08$ & $.11$ & $.39$ & $.39$ 
				& $.66$ & $.06$ & $.06$ & $.75$ & $.40$ & $.38$ \\ 
				means & $\textrm{CI}(\hat{\bar Y}_{HT}^{(H)})$ & $.00$ & $.04$ & $.04$ & $.00$ & $.13$ & $.13$ 
				& $.13$ & $.03$ & $.03$ & $.86$ & $.14$ & $.13$ \\
				\hline				
				\multicolumn{14}{l}{Notes: Results on confidence intervals derived under the assumption of heterogeneous link} \\ 
				\multicolumn{14}{l}{probabilities are based on 500 samples. In Population II the percentages of replicated sam-} \\					
				\multicolumn{14}{l}{ples in which the CIs were not computed because of convergence problems were 0.8\%,} \\
				\multicolumn{14}{l}{8.2\% and 8.8\% in $U_1$, $U_2$ and $U$, respectively, whereas in Population I the respective per-} \\
				\multicolumn{14}{l}{centages were all 0\%. Results on CIs derived under the homogeneity assumption are} \\
				\multicolumn{14}{l}{based on 5000 replicated samples and no convergence problems were presented.} \\				
				%			\end{tabular*}
			\end{tabular}		
		\end{center}
\end{table}}

\renewcommand{\baselinestretch}{1.0} 
\small\normalsize

In the case of the Population II, the performance of the UMLEs $\hat\tau_1^{(U)}$ and $\hat\tau^{(U)}$ was still acceptable, but less good than
that observed in Population I. The performance of $\hat\tau_2^{(U)}$ was not good: it presented a large negative bias and a large variability.
Notice that the distribution of $\hat\tau_1^{(U)}$ was symmetrical and those of $\hat\tau_2^{(U)}$ and $\hat\tau^{(U)}$ were skewed to the right
with long tails. The performance of the HTLEs $\hat\tau_{HT.1}^{(U)}$, $\hat\tau_{HT.2}^{(U)}$ and $\hat\tau_{HT}^{(U)}$ was not good: their
biases were large and caused large values of their $\sqrt{\textrm{r-mse}}$; in fact, the values of both r-bias and $\sqrt{\textrm{r-mse}}$ were
greater than those observed in Population I. The shape of their distributions were similar to those of the corresponding estimators 
$\hat\tau_1^{(U)}$, $\hat\tau_2^{(U)}$ and $\hat\tau^{(U)}$. The performance of the HTLEs $\hat Y_{HT.1}^{(U)}$, $\hat Y_{HT.2}^{(U)}$ and 
$\hat Y_{HT}^{(U)}$ of the population totals was not good in both cases continuous and binary response variables. It is worth noting that their 
distributions were similar to those of the HTLEs $\hat\tau_{HT.1}^{(U)}$, $\hat\tau_{HT.2}^{(U)}$ and $\hat\tau_{HT}^{(U)}$. The performance of 
the HTLEs $\hat{\bar Y}_{HT.1}^{(U)}$, $\hat{\bar Y}_{HT.2}^{(U)}$ and $\hat{\bar Y}_{HT}^{(U)}$ of the population means was acceptable: their 
biases and variances were not large. In the case of the continuous response variable the distribution of $\hat{\bar Y}_{HT.1}^{(U)}$ was skewed 
to the right, whereas those of $\hat{\bar Y}_{HT.2}^{(U)}$ and $\hat{\bar Y}_{HT}^{(U)}$ were skewed to the left. In the case of the binary 
response variable, the distribution of $\hat{\bar Y}_{HT.1}^{(U)}$ was skewed to the right, that of $\hat{\bar Y}_{HT.2}^{(U)}$ was more or less 
symmetrical, and that of $\hat{\bar Y}_{HT}^{(U)}$ was skewed to the left. The performance of the HKLEs of the population totals and means was 
pretty acceptable in both cases continuous and binary response variables, except that of the estimator of $Y_2$ which presented problems of 
subestimation. The distributions of $\hat Y_{HK.1}^{(U)}$, $\hat{\bar Y}_{HK.1}^{(U)}$ and $\hat{\bar Y}_{HK.2}^{(U)}$ were symmetrical, those 
of $\hat Y_{HK.2}^{(U)}$ and $\hat Y_{HK}^{(U)}$ were skewed to the right, whereas that of $\hat{\bar Y}_{HK}^{(U)}$ was slightly skewed to the 
left in the case of the continuous response variable and symmetrical and the case of the binay variable. As in the previous study, the better 
performance of the HKLEs than that of the HTLEs in the case of Population II is result of the weaker correlations between the response variables 
and the inclusion probabilities. The performance of the estimators of the population sizes and population totals derived under the assumption of 
homogeneous link probabilities was bad: they presented serious problems of underestimation. However, the performance of the estimators 
$\hat{\bar Y}_{HT.1}^{(H)}$, $\hat{\bar Y}_{HT.2}^{(H)}$ and $\hat{\bar Y}_{HT}^{(H)}$ was pretty acceptable in both cases continuous and binary 
response variables: they presented small biases and small variances, and in addition, their distributions were more or less symmetrical. We consider 
that the small bias of each one of these estimators could be explained because the bias of each estimator of the population total was practically 
the same as the bias of the corresponding estimator of the population size, and consequently, their biases were canceled out when the quotient was 
computed.

In Table 5 we present the results of the study on the standard deviation estimators. As can be seen, in general the results are not good, but 
there are some exceptions. In Population I, every one of the bootstrap estimators of the standard deviations of the point estimators that were 
derived under the heterogeneity assumption presented a large positive bias, except the estimators of the standard deviations of some estimators 
of parameters of the population $U_1$, that is, $\tau_1$, $Y_1$ and $\bar Y_1$, and those of the estimators $\hat{\bar Y}_{HK.1}^{(U)}$, 
$\hat{\bar Y}_{HK.2}^{(U)}$ and $\hat{\bar Y}_{HK}^{(U)}$ which showed moderate values of bias and variability. The standard deviation estimators 
corresponding to the point estimators derived under the homogeneity assumption showed a large negative bias. In the case of Population II, and 
regardless of the type of response variable, continuous or binary, each one of the standard deviation estimators, either that derived under the 
homogeneity assumption as well as that not derived under that assumption, presented a large negative bias and a large variance. The exceptions 
were again the estimators $\hat{\bar Y}_{HK.1}^{(U)}$, $\hat{\bar Y}_{HK.2}^{(U)}$ and $\hat{\bar Y}_{HK}^{(U)}$, which in the case of the binary 
response variable presented small values of bias and moderate values of variability.

The results of the Monte Carlo study on the confidence intervals are shown in Table 6. In the case of  Population I, and regardless of the type
of response variable, continuous or binary, every one of the CIs that was derived under the heterogeneity assumption presented good coverage
probabilities, except the CIs obtained from the HTLEs $\hat\tau_{HT.1}^{(U)}$, $\hat\tau_{HT.2}^{(U)}$ and $\hat\tau_{HT}^{(U)}$, and from the
HKLEs $\hat Y_{HK.1}^{(U)}$ and $\hat Y_{HK}^{(U)}$, which presented relatively low values of the cp, and those obtained from the HKLEs of the 
population means which showed very low values of the cp. Notice that in the case of the continuous response variable the CIs obtained from these 
estimators presented null values of the coverage probabilities. On the other hand, with respect to the relative lengths of the CIs, we have that 
the CIs of the parameters associated with $U_2$, that is, $\tau_2$, $Y_2$ and $\bar Y_2$, showed very large lengths, and consequently their 
performance was not good. The exception was the CI obtained from the HKLE $\hat{\bar Y}_{HK.2}^{(U)}$, which showed a small value of mdrl. The 
relative lengths of the CIs of $\tau_1$, $Y_1$ and $\bar Y_1$, as well as those of $\tau$, $Y$ and $\bar Y$ were acceptable, and consequently their 
performance was also acceptable. The CIs derived under the homogeneity assumption presented null values of their cp, which were consequence of the 
large biases of the corresponding point estimators. Therefore, despite the adequate values of their relative lengths, their performance was bad. 
In the case of Population II, the cp of the CIs that were constructed under the assumption of heterogeneous link probabilities were in general small. 
Only those associated with the CIs for $\bar Y_2$ (for both types of response variables), and particularly those corresponding to the CIs of the means 
based on the HKLEs were acceptable, say greater than 0.9. The small values of the cp were consequence of the biases of the corresponding point estimators. 
The relative lengths of these CIs were acceptable. Thus, based on the values of the cp and relative lengths, only the CIs of the population means of the 
binary variable obtained from the HKLEs had good performance, as well as those of mean $\bar Y_2$ of both variables (continuous and binary) obtained from 
the HTLE $\hat{\bar Y}_{HT.2}$. The CIs of $\tau_1$ based on $\hat\tau_1^{(U)}$ and those of the means of the binary variable based on the HTLEs showed 
regular performance. It is worth noting that even though the CIs of the means based on the HTLEs showed low values of the cp, the average CIs of the means 
$\bar Y_1=49.99$, $\bar Y_2=39.87$ and $\bar Y=47.47$ of the continuous variable were $(42.40,45.90)$, $(34.29,41.47)$ and $(42.72,46.52)$, respectively, 
whereas the average CIs of the means $\bar Y_1=49.91$, $\bar Y_2=40.28$ and $\bar Y=47.51$ of the binary variable were $(41.43,50.41)$, $(31.22,51.76)$ 
and $(40.15,49.17)$, respectively; therefore, these CIs still provide valuable information. Finally, the CIs constructed under the assumption of homogeneous 
link probabilities also presented small values of their cp, and consequently their performance was not good. The only exception were the CIs of the population 
means of the binary response variable where their cp showed moderate values, that is, between 0.75 and 0.91. Thus, based on the values of their cp and mdrl 
these CIS had regular performance. However, the average CIs of the means of the continuous variable were $(50.24,51.74)$, $(39.47,41.86)$ and 
$(48.18,49.61)$, whereas those of the means of the binary variable were $(46.17,52.66)$, $(37.46,52.77)$ and $(45.37,51.55)$; consequently, the information 
provided by these CIS provide is useful.

From the results of this study we have that inferences based on the UMLEs of the population totals are pretty acceptable, although the coverage probabilities
of the CIs are affected when the assumed model of the link probabilities is not satisfied. Inferences based on the HTLEs of the population sizes are not good. 
HTLEs of the totals and means perform well when the correlation between the values of the response variable and the inclusion probabilities is not too small, 
say greater than or equal to $0.3$. For values of the correlation coefficient less than $0.3$, the HTLEs of the means are the only ones that perform acceptably,
although the values of the coverage probabilities of the CIs are affected. HKLEs of the totals and means tend to perform acceptably when the values of the 
correlation coefficient between the response variable and the inclusion probabilities are small. In particular, HKLEs of the means perform pretty well in this
situation.

\section{Conclusions and suggestions for future research}

In this work we have considered the link-tracing sampling variant proposed by F\'elix-Medina and Thompson (2004) and have proposed Horvitz-Thompson-like 
and H\'ajek-like estimators of population totals and means. This work extends that of F\'elix-Medina and Monjadin (2010) by assuming heterogeneous, rather
than homogeneous, link probabilities which are modeled by a Rasch model used by F\'elix-Medina et al. (2015). The variances of the proposed estimators are
estimated by a variant of bootstrap which extends the variant used by the previously cited authors by estimating the variances of estimators of totals and
means, in addition to the variances of estimators of population sizes. This variant of bootstrap allows the estimation of variances when the response variable
is either continuous, discrete or binary. Large sample confidence intervals for the population parameters are constructing by assuming normal distributions 
of the estimators of the parameters, except in the case of the estimators of the population sizes, where log-normal distributions are assumed. In addition, 
confidence intervals for proportions are constructed busing Korn and Graubard's (1998) proposal.

We evaluated the performance of the proposed estimators by means of two Monte Carlo studies. In the fist study a finite population was constructed using data 
from the Add Health study. The result of this numerical study are promising; thus, if this population were more or less representative of the populations that 
could be found in applications of this methodology, then reliable inferences would be expected to be obtained. However, the results of the second study in
which two populations were constructed using simulated data show that erroneous inferences might be obtained if some model assumptions were not satisfied. In
particular, we found that if any of the assumptions is satisfied, then reliable inferences about population sizes, totals and means are obtained. Furthermore,
we found that the assumption of the Poisson distribution of the sizes $M_i$s of the venues $A_i$s does not need to be satisfied to obtain reliable inferences.
Nevertheless, we also found out that severe deviations from the Rasch model of the link probabilities lead to erroneous inferences, and that inferences
about population sizes and totals are affected in greater extent than inferences about population means. In fact, inferences about the population means seem
to be robust to deviations from the assumed models. In addition, we came upon that in any situation, the performance of the proposed bootstrap variance 
estimators is at most just good.

In the light of these results, we consider that the following issues are worthy of future research: (i) To develop a model for the link probabilities 
$p_{ij}^{(k)}$s that be robust to deviations from the assumed model. For instance, to model $p_{ij}^{(k)}$ as a quadratic function of $\alpha_i^{(k)}$ and 
$\beta_j^{(k)}$, or to assume that the $\beta_j^{(k)}$s are T-Student distributed instead of normally distributed, or to change the Rasch model by the latent
classes model proposed by Pledger (2000) and which was used in the second Monte Carlo study to generate the values $x_{ij}^{(k)}$s in Population II. (ii) To
improve the proposed bootstrap variance estimators. For example, to predict the values of the response variable associated with the nonsampled elements by
using a quadratic or a nonparametric regression model instead of a simple linear regression model. (iii) To enhance the proposed CIs of the population sizes,
totals and means. For instance, using the bootstrap percentile method which does not require assuming a probability distribution for the estimator of the 
parameter of interest.

\section*{Acknowledgements}

This research was supported by Grant PROFAPI-2011/057 of Universidad Aut\'onoma de Sinaloa.

\nocite{AbramowitzStegun64}
\nocite{Bernardetal10}
\nocite{Boothetal94}
\nocite{Chao87}
\nocite{Chao88}
\nocite{Chengetal20}
\nocite{ChowThompson03}
\nocite{CoullAgresti99}
\nocite{Crawfordetal18}
\nocite{DavisonHinkley97}
\nocite{Dombrowskietal12}
\nocite{FelixThompson04}
\nocite{FelixMonjardin06}
\nocite{FelixMonjardin10}
\nocite{Felixetal15}
\nocite{FrankSnijders94}
\nocite{GinerSmyth16}
\nocite{Harris13}
\nocite{Handcock14}
\nocite{Heckathorn97}
\nocite{Heckathorn02}
\nocite{HeckathornCameron17}
\nocite{JohnstonSabin10}
\nocite{Johnstonetal13}
\nocite{Johnstonetal16}
\nocite{Kalton09}
\nocite{Khanetal18}
\nocite{Killworthetal98a}
\nocite{Killworthetal98b}
\nocite{Klovdahl89}
\nocite{KornGraubard98}
\nocite{Lee-etal14}
\nocite{Magnanietal05}
\nocite{Maltiel15}
\nocite{MarpsatRazafindratsima10}
\nocite{MacKellaretal96}
\nocite{McCormicketal10}
\nocite{MengGustafson17}
\nocite{Pledger00}
\nocite{Rsystem18}
\nocite{Sanathanan72}
\nocite{Spreen92}
\nocite{SpreenBogaerts15}
\nocite{Tourangeau14}
\nocite{ThompsonFrank00}
\nocite{Thompson02}
\nocite{StaudteSheather90}
\nocite{StClairOConnell12}
\nocite{UNAIDS-WHO10}
\nocite{VolzHeckathorn08}
\nocite{Williams02}
\nocite{Zelterman88}

\bibliographystyle{achicago}
\bibliography{EstTotalsMeansLTS-HK-rt}

\end{document}